\documentclass[letterpaper,10pt,journal,final]{IEEEtran}
                      \pdfoutput=1

\IEEEoverridecommandlockouts                            

\usepackage[eulergreek]{sansmath}
\usepackage[utf8]{inputenc}
\usepackage{float, amsmath, amssymb, graphicx, color, tikz,tabularx,algorithm}
\usepackage{tikzscale,pgfplots, pdftexcmds, enumerate,algpseudocode}
\usepackage{mathtools,url}
\usepackage[noadjust]{cite}
\usepackage[capitalise]{cleveref}

\pgfplotsset{width = 0.48\textwidth, compat = 1.13, height = 7cm, grid= major, 
			ticklabel style = {font=\sansmath\sffamily\scriptsize},
			ylabel style = {font = \sf\footnotesize, yshift = -0.2cm},
			xlabel style = {font = \sf\footnotesize, yshift = 0.15cm},
			zlabel style = {font = \sf\footnotesize},
			legend style = {font=\sf\scriptsize}, legend cell align = left, 
			title style={yshift=-7pt, font =\sf\footnotesize} }
\usepgfplotslibrary{fillbetween,groupplots}
\usetikzlibrary{arrows,shapes,backgrounds,patterns,fadings,decorations,
				positioning,external,arrows.meta}
\tikzset{
	box/.style={rectangle, thin, inner sep = 6pt, minimum height=2.4em, text 
	centered, draw=black},
	node_rec/.style={rectangle, thin, inner sep = 3pt, minimum size=2em, text 
		centered, draw=black},
	node_cir/.style={circle,thin,inner sep = 0pt,minimum size=0.5em,text 
		centered,draw = black},
	circ/.style={circle,thin,inner sep=0pt,minimum height=.8em,text	centered, 
		draw = black},
	rec/.style={rectangle,thin,inner sep=0pt,minimum height=.8em},
	connection/.style={->,semithick, shorten <=0pt, shorten >=1pt},
	node/.style={circle, thin, inner sep = 0pt, minimum height=2em, text 
	centered, draw=black},
	dot/.style={circle,fill=black,minimum size=4pt,inner sep=0pt,outer sep=-1pt}
}
\graphicspath{{./pic/},{./fig/}}

\title{\LARGE \bf  Feedback Linearization based on Gaussian \\ 	
Processes with event-triggered Online Learning}

\binoppenalty=\maxdimen 
\relpenalty=\maxdimen   

\author{Jonas~Umlauft,~\IEEEmembership{Student~Member,~IEEE},
	    Sandra~Hirche,~\IEEEmembership{Senior~Member,~IEEE}%
	\thanks{All authors are members of the Chair of Information-oriented 	
			Control, Department of Electrical and Computer Engineering,
			Technical University of Munich, D-80333 Munich, Germany,
			 \mbox{fax: +498928928340}, \mbox{telephone: +498928923403},
		{\tt\small [jonas.umlauft, hirche]@tum.de}}%
	\thanks{The research leading to these results has received funding
		from the European Research Council under the European
		Union Seventh Framework Program (FP7/2007-2013) / ERC
		Starting Grant ``Control based on Human Models (conhumo)'' agreement 
		no. 337654.}%
	\thanks{The work has been submitted to the IEEE Transactions on Automatic 
	Control on Dec 10th, 2018. }%
}

\renewcommand\vec\boldsymbol 

\DeclarePairedDelimiter{\norm}{\lVert}{\rVert}

\newcommand{\smin}{\sigma_\text{min}}
\newcommand{\smax}{\sigma_\text{max}}
\newcommand{\T}{^\intercal}
\newcommand{\N}{\mathcal{N}}
\newcommand{\Nset}{\mathbb{N}}
\renewcommand{\O}{\mathcal{O}}
\newcommand{\Rset}{\mathbb{R}}
\newcommand{\x}{\vec{x}}
\newcommand{\Xset}{\mathbb{X}}

\newcommand{\alli}{\mbox{$i=1,\ldots,N$ }}
\newcommand{\allj}{\mbox{$j=1,\ldots,n$ }}
\newcommand{\allx}{\mbox{$\forall \x \in \Xset$}}
\newcommand{\allxt}{\mbox{$\forall \x \in \Xsett$ }}

\newcommand{\bpsi}{\vec{\psi}}

\newcommand{\Bset}{\mathbb{B}}
\newcommand{\ckb}{\vec{\lambda}}
\newcommand{\ck}{\lambda}
\newcommand{\e}{\vec{e}}

\newcommand{\dt}{\Delta t}
\newcommand{\df}{\Delta \f}
\newcommand{\diag}{\text{diag}}
\newcommand{\Dset}{\mathbb{D}}
\newcommand{\Dk}{\Dset_{\kap}}
\newcommand{\E}{\mathbb{E}}
\newcommand{\f}{f}
\newcommand{\fx}{\f(\x)}
\newcommand{\fdot}{\f(\cdot)}
\newcommand{\fxi}{\f\left(\xhiv\right)}
\newcommand{\fh}{\hat{\f}}

\newcommand{\fhk}{\fh_\kap}
\newcommand{\fhkx}{\fh_\kap(\x)}
\newcommand{\fhkdot}{\fh_\kap(\cdot)}
\newcommand{\fhx}{\fh(\x)}

\newcommand{\ftrue}{\f_{\text{true}}}
\newcommand{\fsum}{f_\text{sum}}
\newcommand{\fprod}{f_\text{prod}}
\newcommand{\fGP}{f_{\GP}}
\newcommand{\g}{g}
\newcommand{\gx}{\g(\x)}
\newcommand{\gdot}{\g(\cdot)}
\newcommand{\gxi}{\g\left(\xhiv\right)}
\newcommand{\gh}{\hat{\g}}
\newcommand{\ghx}{\gh(\x)}
\newcommand{\ghk}{\gh_\kap}
\newcommand{\ghkx}{\gh_\kap(\x)}
\newcommand{\ghkdot}{\gh_\kap(\cdot)}
\newcommand{\gb}{\bar{\g}}
\newcommand{\GP}{\mathcal{GP}}
\newcommand{\kc}{k_c}
\newcommand{\kap}{\kappa}
\newcommand{\K}{\vec{K}}
\renewcommand{\k}{\vec{k}}
\newcommand{\I}{\vec{I}}

\newcommand{\Lsig}{L_{\sigma}}

\newcommand{\mX}{\vec{m}^{\vec{X}}}
\newcommand{\sigon}{\sigma_{\text{on}}^2}
\newcommand{\sigonsd}{\sigma_{\text{on}}}

\newcommand{\tk}{t_{\kap}}
\newcommand{\tkp}{t_{\kap+1}}
\newcommand{\tlb}{t_\text{lb}}

\newcommand{\Uo}{\vec{U}}
\newcommand{\Uset}{\mathbb{U}}
\newcommand{\V}{\mathbb{V}}

\newcommand{\xhiv}{\x^{(i)}}

\newcommand{\Xsett}{\tilde{\Xset}}

\newcommand{\y}{\vec{y}}
\newcommand{\yd}{\rho}
\newcommand{\yhi}{y^{(i)}}
\newcommand{\ybsum}{\vec{y}_{\text{sum}}}
\newcommand{\ysum}{y_{\text{sum}}}
\newcommand{\ybprod}{\vec{y}_{\text{prod}}}
\newcommand{\yprod}{y_{\text{prod}}}

\crefname{definition}{Definition}{Definitions}

\newtheorem{thm}{Theorem} 
\crefname{thm}{Theorem}{Theorems}

\newtheorem{assum}{Assumption} 
\crefname{assum}{Assumption}{Assumptions}

\newtheorem{cor}{Corollary} 
\crefname{cor}{Corollary}{Corollaries}

\newtheorem{lem}{Lemma}
\crefname{lem}{Lemma}{Lemmata}

\newtheorem{prop}{Proposition}
\crefname{prop}{Proposition}{Propositions}

\crefname{problem}{Problem}{Problems}

\crefname{property}{Property}{Properties}

\newtheorem{remark}{Remark}
\crefname{remark}{Remark}{Remarks}

\crefname{section}{Sec.}{Sec.}
\Crefname{section}{Section}{Sections}

\colorlet{diffc}{black}

\begin{document}

\maketitle
\thispagestyle{empty}
\pagestyle{empty}

\begin{abstract}
Combining control engineering with nonparametric modeling techniques 
from machine learning allows to control systems without analytic description 
using data-driven models. Most existing approaches separate 
\textit{learning}, i.e. the system identification based on a fixed 
dataset, and \textit{control}, i.e. the execution of the model-based 
control law. This separation makes the performance highly sensitive to the 
initial selection of training data and possibly requires very 
large datasets. This article proposes a learning
feedback linearizing control law using online closed-loop 
identification. The employed Gaussian process model updates its training data 
only if the model uncertainty 
becomes too large. This event-triggered online learning ensures 
high data efficiency and thereby reduces the 
computational complexity, which is a major barrier for using  Gaussian 
processes under real-time constraints. We propose safe forgetting
strategies of data points to adhere to budget constraint and to 
further increase data-efficiency. We show asymptotic stability for the tracking 
error under the proposed event-triggering law and illustrate the effective 
identification and control in simulation.
\end{abstract}

\section{Introduction}
\label{sec:intro}

\IEEEPARstart{D}{ata-driven} control gained high attention as costs for 
measuring, processing 
and storing data is rapidly decreasing and control engineering is 
increasingly applied in areas where it is difficult to describe the plant using 
first principles. Nevertheless, a precise system description is essential for 
many modern model-based control algorithms as e.g. model 
predictive control and feedback linearization. Classical system identification 
using parametric models like autoregressive moving average (ARMA) or 
Hammerstein models~\cite{ljung1998system}, reaches its limits when the choice 
of a suitable model class is cumbersome or impossible, e.g. in systems where 
human behavior is part of the control loop. That is where 
data-driven nonparametric models have their advantages as only 
minimal prior knowledge is required and allow higher flexibility than 
parametric models.

This article particularly considers Gaussian processes (GPs) which are well 
recognized in machine learning and control for modeling complex 
dynamics~\cite{kocijan2016modelling}. The Bayesian background allows an 
implicit bias-variance trade-off~\cite{rasmussen2006gaussian} and as GPs are a 
kernel based method, prior knowledge (if any exists) can properly be 
transferred 
into the model~\cite{duvenaud2014thesis}. The major advantage is, that GP 
models also encode their own ignorance and therefore provide information 
whether the model is reliable for particular inputs or not.

Due to their nonparametric nature, the model complexity of a GP increases with 
the number of available data. This possible unlimited expressive power is 
generally desired, however may cause difficulties from a computational point of 
view  in the case of large training data sets. Particularly challenging are 
online learning 
schemes where data points are accumulated over time and real-time capability is 
critical. This raises the question of efficient online learning
strategies 
for 
nonparametric models. Time-triggered model adaptation fails to 
distinguish 
whether a new measurement or training point is necessary at the current 
location of the state-space or not. This calls for an event-triggered scheme, 
which decides upon a new measurement based on the current reliability of the 
model, which is expected to result in higher data-efficiency.

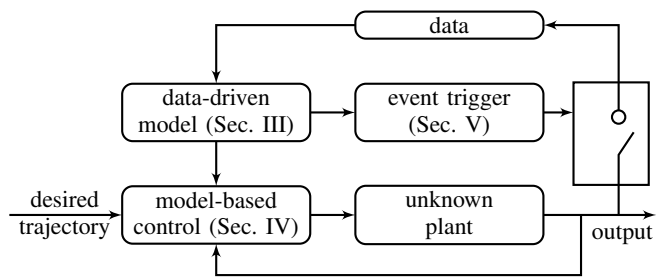
\begin{figure}
	\vspace{1.5mm}
	\tikzstyle{block}=[draw, inner sep=2, outer sep=0, minimum width=2em, 
	rounded corners, thick,  align=center]
	\centering  \tikzsetnextfilename{ctrl_struct_eventonline}
	\tikzset{external/export next=false}
	\begin{tikzpicture}[node distance=2cm, 	auto,>=latex',
						every text node part/.style={align=center},
						thick,scale=1, every node/.style={scale=1,font=\small}]
	\def\dis{0.6}
	\def\sh{0.1}
	\def\mw{2.5cm}
	\node [block,minimum width=\mw] (plant) {unknown \\  plant };
	\node [block,left = \dis of plant,minimum width=\mw] (ctrl) 
					{model-based \\control (\cref{sec:control})};
	\node [block,above =\dis  of ctrl,minimum width=\mw] (model) 
					{data-driven\\ model (\cref{sec:id})};	
	\node [block,right = \dis of model,minimum width=\mw] (event) 
					{event trigger \\ (\cref{sec:event})};
	\node [block,above =\dis  of event,minimum width=\mw] (data) {data};	
	\draw[->] (model.south) -- (ctrl.north);
	\draw[->] (ctrl.east) -- (plant.west); 
	\draw[<-] (model.north) |-	(data.west) ;

	\draw[->] (model.east)-- (event.west);

	\draw[->] ([shift = {(-1.5,0)}]ctrl.west)--(ctrl.west) 
				node[pos=0, anchor = west] {\small desired \\ \small 
				trajectory};
	\draw[->] (plant.east) -- ([shift = {(1.5,0)}]plant.east)
			node[pos=0.7,anchor = north] {\small output};
	\draw[->]  ([shift = {(0.5,0)}]plant.east)  --++ (0,-0.8)-| 
	(ctrl.south);		
	\draw[-]  ([shift = {(1,0)}]plant.east)  --++ (0,0.8)--++ (0.2,0.3);
	\draw[->] (event.east) -- ([shift = {(0.4,0)}]event.east);
	\draw[<-o] (data.east) -| ([shift = {(1,1.2)}]plant.east);
	\draw ([shift = {(0.4,0.4)}]event.east)  rectangle 
				([shift = {(1.4,0.4)}]plant.east);
		
	\end{tikzpicture}
	\caption{Proposed concept of an online learning control law 
	with event-triggered model updates.}
	\label{fig:concept}
\end{figure}
\subsection{Related work} 
The fact that no model can initially capture all aspects of the true system 
motivated robust and adaptive control methods to overcome this 
discrepancy~\cite{ioannou1996robust}. The online adaptation of the control 
strategy or its employed model is well understood for parametric 
models~\cite{krstic1995nonlinear},~\cite{aastrom2013adaptive}. {\color{diffc} 
In particular for linear systems, data-driven 
approaches are extensively researched, 
see~\cite{bazanella2011data}, and~\cite{silva2019data}. For nonlinear systems, }
model 
reference adaptive control (MRAC) is designed to effectively deal with model 
uncertainties or only little prior knowledge using online parameter 
estimation{\color{diffc}~\cite{campestrini2017data}}. 
Iterative learning control (ILC) improves control performance by iteratively 
modulating the control signal in a repetitive task, such that experience from 
earlier executions are used to improve performance, 
{\color{diffc}see~\cite{bristow2006survey} and~\cite{radac2014iterative}.} 
However, most existing MRAC and ILC methods are mainly based on parametric 
models which suffer from limited complexity and flexibility. 
{\color{diffc}Model-free adaptive control (MFAC) avoids an explicit model but 
instead employs e.g. a dynamic linearization~\cite{hou2017overview}, virtual 
reference feedback tuning~\cite{campi2006direct} or a closed-loop control 
parameter optimization~\cite{hjalmarsson1998iterative}. 
Alternatively, a spectral analysis for nonparametric frequency-domain tuning is 
considered~\cite{kammer2000direct} or extremum seeking is 
employed for performance optimization~\cite{killingsworth2006pid}. }

Other than in classical control theory, the machine learning literature employs 
more frequently data-driven models with infinite expressive power for online 
adaptation~\cite{theodorou2010reinforcement}. The class of  model-based 
reinforcement learning algorithms considers continuous model and controller 
updates to maximize a reward~\cite{sutton1998reinforcement}. For 
example~\cite{deisenroth2011pilco} shows a high data efficiency with Gaussian 
process models. These are also successfully applied in 
robotics~\cite{nguyen2011model,nguyen2009model}, however most approaches 
miss a formal stability analysis for the system's behavior. 

Very recently, several control approaches with formal guarantees for GP models 
have been developed
which, however, keep a fixed dataset during execution of the control 
law~\cite{beckers2017stable},~\cite{fanger2016gaussian}. The work 
in~\cite{beckers2017bstable} considers the control of 
Lagrangian systems and shows boundedness of the tracking error. The 
identification of a priori known stable systems with GPs is analyzed 
in~\cite{umlauft2017learning}, and~\cite{umlauft2018uncertainty} proposes an 
uncertainty-based control approach for which asymptotic stability is proven. 
However, none of these techniques updates the model while controlling the 
system. The work in~\cite{berkenkamp2016safe} proposes a safe exploration by 
sequentially adding training points to the dataset, but it only 
stays within the region of attraction and cannot track an arbitrary trajectory 
in the state space.

An online learning tracking control law with with time-triggered 
adaptation is proposed in~\cite{chowdhary2015bayesian}. As a result data points 
are added to the training dataset irrespectively of their importance. This 
might compromise real-time capability as the computational inefficiency for 
large datasets is a known challenge of GPs~\cite{rasmussen2006gaussian}.

This difficulty in circumvented 
in~\cite{chakrabortty2009robust,chakrabortty2009time}, where the unknown 
dynamics 
is estimated using high gain filters. However, these 
approaches suffer from the known difficulties of high gain control, i.e. a 
quick saturation of input signals and the amplification of noise. The latter is 
avoidable by combining feedback and 
model-based feedforward control.
This idea is not just employed in this article, but was also used 
in~\cite{polycarpou1991identification}, where a neural network identifies the 
dynamics without any parametric prior knowledge.
Particularly~\cite{lewis1996multilayer},~\cite{lewis1995neural} 
and~\cite{sanner1991stable} focus on stability and performance guarantees. The 
work in~\cite{yecsildirek1995feedback} proposes a feedback linearizing control 
law, which adapts online the weights of a neural network model. It 
shows boundedness of the adaptation law and the resulting controller but cannot 
quantify the ultimate bound because neural networks - in comparison to 
GPs - do not inherently provide a measure for the fidelity of the 
model~\cite{umlauft2017bayesian}. This becomes important, if the 
controller is applied in safety critical domains, where the tracking 
error must be quantified to avoid failure or damage to the system.

In summary, to date there exists no approach, which adapts a 
nonparametric model online to guarantee asymptotic stability of the tracking 
error. Thus for universal models, which can represent arbitrarily 
complex 
dynamics, there are missing online learning control laws, to guarantee safe 
behavior of the 
closed-loop system. Also a data-efficient update strategy is required to keep 
the model computationally efficient, which is important in many real-time 
critical applications.

\subsection{Contribution and structure}
The main contribution of this article is an online learning 
feedback linearizing control law based on Gaussian processes for an initially 
unknown system. This control algorithm includes a closed-loop 
identification scheme for control affine systems 
exploiting compound kernels for GPs. To ensure data-efficiency of 
the approach, we propose an event-triggered online learning mechanism which 
decides upon a model update based on its current reliability. The derivation is 
based on a probabilistic upper bound for the model error of a GP, and allows to 
provide safety guarantees in terms of convergence properties of the closed-loop 
system. For noiseless training data,
we show global asymptotic stability and for noisy output training 
data global 
ultimate 
boundedness of the tracking error. For the case of a constraint budget for 
data points, we propose a forgetting strategy, which maintains the  
convergence guarantees using a reduced number of training points.

The article is based on the preliminary work in~\cite{umlauft2017feedback}, 
which focuses on the identification of a control affine system with GPs given a 
fixed dataset. In contrast, this work considers the online collection of data 
and updates the model while the control law is active. This allows to show 
asymptotic stability with a data-efficient event-triggered update rule
while~\cite{umlauft2017feedback} only showed existence of an ultimate bound. 

This article is structured as follows:
After formulating the considered problem formally in \cref{sec:problem}, 
\cref{sec:id} reviews the identification of control affine systems based on GPs.
In \cref{sec:control}, the feedback linearizing tracking control law is 
proposed including a convergence analysis for training data 
measured online at arbitrary time instances. \Cref{sec:event} introduces an 
event-triggering  to  
update the model based on its uncertainty. A numerical illustration is 
provided in \cref{sec:sim} followed by a conclusion in \cref{sec:conclusion}.

\subsection{Notation}
\label{sec:notation}
Lower/upper case bold symbols denote vectors/matrices, 
$\Rset_{+,0}$/$\Rset_+$ all real positive numbers with/without zero, 
$\Nset_{0}$/$\Nset$ all natural numbers with/without zero, 
$\smin(\cdot),\smax(\cdot)$ the minimal/maximal singular value of a matrix and
$\E[\cdot]$/$\V[\cdot]$ the expected value/variance of a random variable, 
respectively.
$\I_n$ denotes the~$n \times  n$ identity matrix,
$\N(\mu,\sigma)$ a Gaussian distribution with mean~$\mu$ and variance~$\sigma$,
$\vec{a}_{1:n}$ the first~$n$ elements of the vector~$\vec{a}$,
$\cdot \succ 0$ the positive definiteness of matrix or function and 
$\|\cdot\|$ the Euclidean norm if not stated otherwise.

\section{Problem Formulation}
\label{sec:problem}
Consider a single-input system in the controllable canonical form
\begin{align}
\label{eq:sys}
\dot{x}_1 &= x_2 \nonumber\\
\dot{x}_2 &= x_3 \nonumber\\
		  &\ \cdots \nonumber\\ 
\dot{x}_n &= \fx + \gx u, \quad \x_0 = \x(0),
\end{align}
with state~$\x =[x_1\ x_2 \cdots x_n]\T \in \Xset \subseteq \Rset^n$ and 
input~$u \in \Uset = \Rset$; the functions~$\fdot$ and~$\gdot$ 
are considered unknown. The following assumptions are made.
\begin{assum}
	\label{assum:fgbounded}
	The unknown functions \mbox{$\f:\Xset \to \Rset$} and~\mbox{$\g:\Xset \to 
	\Rset$} are globally bounded and differentiable. 
\end{assum}
Differentiability is a very natural assumption, as it holds for many 
physical systems. The boundedness of the functions~$\fdot,\gdot$ would  
automatically be implied (due to the differentiability) if the set~$\Xset$ was 
bounded. However, we want~$\Xset$ to be possibly unbounded. 

From \cref{assum:fgbounded}, the first property is derived.
\begin{lem}
	\label{lem:finiteEscapetime}
	Consider the system~\eqref{eq:sys} under \cref{assum:fgbounded} with 
	bounded and continuous~$u(\x)$. Then the solution~$\x(t)$ does not have a 
	finite escape time, thus~$\nexists t_{\infty}$,~$0<t_{\infty}<\infty$ for 
	which
	\begin{align}
		\lim\limits_{t\to t_{\infty}}\norm{\x(t)}=\infty.
	\end{align}
\end{lem}
\begin{IEEEproof}
	According to \cite[Theorem 3.2]{khalil1996nonlinear} the stated conditions 
	ensure a unique solution~$\x(t)$, for all~$t >0$ for which the finite 
	escape time follows from the differentiability of~$\fdot,\gdot$ and 
	the bounded control input.
\end{IEEEproof}
As a stabilizing controller is not known in advance (because~$\fdot,\gdot$ are 
unknown), the absence of a finite escape time  is important: It allows to 
collect observations of the system in any finite time interval with a ``poor'' 
controller (or also~$u(\x)=0$) without risking damage due to ``infinite'' 
states. {\color{diffc}Additionally, we assume the following.
\begin{assum}
	\label{assum:degree_signg}
	For system~\eqref{eq:sys} holds~$\gx > 0, \allx$.
\end{assum}
This ensures that the system's relative degree is equal to the system order~$n$ 
for all~$\x \in \Xset$ and the sign of~$\gdot$ is known. 
Equivalently,~$\gdot$ can also be taken as strictly negative resulting 
in a change of sign for the control input.}
\Cref{assum:degree_signg} is necessary to ensure global controllability and 
excludes the existence of internal dynamics. It restricts the system class, 
however the focus on this work is on the online learning control and 
extending it to a larger system classes is part of future work.

We assume that observations are taken online while the proposed control law is 
active.
\begin{assum}
	\label{assum:TrainingData}
	Noiseless measurements of the state vector~$\x^{(\kap)} = 
	\x(\tk)$ and 
	noisy 
	measurements of the highest derivative~\mbox{$\y^{(\kap)} = \dot{x}_n(\tk) 
	+ \epsilon^{(\kap)}$} can be taken at arbitrary time instances~$\tk$ 
	with~$\kap \in \Nset_0$. The observation noise~\mbox{$\epsilon^{(\kap)} 
	\sim \N(0,\sigon)$} is assumed Gaussian, independent and identically 
	distributed. The time-varying dataset
	\begin{align}
	\label{eq:def_Dset}
	\Dk = \left\{\xhiv, \yhi \right\}_{i=1}^{N_\kap},
	\end{align}
	is updated at time~$\tk$ and remains constant 
	until~$\tkp$ and~$N_\kap\in \Nset_0$ denotes the current number of data 
	points.
\end{assum}
The exact measurement of the state is a common assumption and necessary for 
feedback linearization. The time derivative of the state~$x_n$ can, for 
practical applications, be approximated through finite differences. The 
approximation error is then considered as part of the measurement noise 
as other additive sources of imprecision result in an overall 
sub-Gaussian noise distribution. 
Alternatively, a separate sensor for measurements of~$\dot{x}_n$ is necessary.

Throughout this article, we will refer to~$\sigon=0$ as the \textit{noiseless 
case} and~$\sigon>0$ as the \textit{noisy case} considering measurements 
of~$\dot{x}_n$. The measurement of the state~$\x$ will always be assumed noise 
free.

Consider, that~$N_\kap$ is not necessarily increasing with increasing~$\kap$ as 
data pairs can also be discarded from the 
dataset if not needed anymore. However, this set~$\Dk$ remains constant 
between 
two consecutive measurements, because elements are only added or removed 
at~$\tk$.

The goal is to design an online learning feedback linearizing 
control 
law - based on dataset~$\Dk$ - of the form
\begin{align}
\label{eq:FeLin}
u_\kap(\x)= \frac{1}{\ghkx}\left(-\fhkx + \nu\right), \quad 
\kap \in\Nset_0,
\end{align}
where~$\nu \in \Rset$ is the input to the resulting approximately linearized 
system and the functions~$\fhk:\Xset\to\Rset$,~$\ghk:\Xset\to\Rset$ are the 
approximations for the unknown functions~$\fx$,~$\gx$. The control 
law~\eqref{eq:FeLin} is switching, because the model~$\fhkx$,~$\ghkx$ 
is updated with every change of the dataset~$\Dk$ at time~$\tk$. We 
would like to emphasize, that measurements are not taken at a constant time 
interval, and updates are therefore not performed periodically. 
Instead, the 
updates will be performed when needed, i.e. triggered by an event 
(introduced in \cref{sec:event}) and 
thus~$\tk$ for~$\kap\in\Nset_0$ are not equidistant. By definition, 
the~$\kap$-th update occurs at~$\tk$ and the control law~$u_\kap$ is then 
applied until the next event at~$\tkp$, more formally written as
\begin{align}
\label{eq:uswitch}
u(\x) = u_\kap(\x),\quad t\in [\tk\ \tkp).
\end{align}

\section{Gaussian Process Learning for Control Affine Systems}
\label{sec:id}
For the closed-loop online identification of~$\fdot$ and~$\gdot$ we consider 
Gaussian process regression, which then provides the 
approximations~$\fhkdot$ and~$\ghkdot$. We will first introduce GP regression 
in general (\cref{sec:GPs}), before presenting our tailored solution for 
control affine closed-loop systems in \cref{sec:GP4ctrlaffinesys}.
\subsection{Gaussian process regression}
\label{sec:GPs}
Consider a function~$\ftrue \colon \Xset \to \Rset$ for which noisy measurements
of the image at the locations~$\xhiv \in \Xset$ are available, thus
\begin{align}
	y_f^{(i)} = \ftrue\left(\xhiv \right) + \epsilon^{(i)},
\end{align}
where \mbox{$\epsilon^{(i)} \sim \N(0,\sigon)$} and~$\alli$ (where we write 
simply~$N$ for~$N_\kap$ in this section). Modeling this function with a 
Gaussian process~$\fGP(\x)$ results in a stochastic process which assigns a 
Gaussian distribution to any finite subset~$\{\x_1,\ldots,\x_M\} \subset \Xset$ 
in a continuous domain. The GP is also often considered as distribution over 
functions~\cite{rasmussen2006gaussian}, denoted by
\begin{align}
\fGP(\x) \sim \GP\left(m(\x),k(\x,\x') \right),
\end{align}
and is fully specified by a mean~$m(\x): \Xset \to \Rset$ and 
covariance~$k(\x,\x'): \Xset \times \Xset \to \Rset$ function.
The mean function includes prior knowledge of the function~$\ftrue$ if there is 
any. Otherwise, it is commonly set to zero. The covariance function, also 
called kernel 
function, determines properties of~$f_{\GP}(\x)$, like the smoothness and 
signal variance. Mean and kernel function are described by the 
hyperparameters~$\bpsi$. 

Using Bayesian techniques, the likelihood function
\begin{align}
\label{eq:opt_like}
\bpsi^*&=\arg\max\limits_{\bpsi} \log p(\y_f |\vec{X},\bpsi),\\
\log p(\y_f |\vec{X},\bpsi)&=
			\frac{1}{2}\left(\y_f^T\K^{-1}\y_f-\log\det\K
				-N\log(2\pi)\right) \nonumber,
\end{align}
is maximized to obtain the optimal hyperparameters for a given set of 
observations. As notation we use 
\begin{align}
\label{eq:defX}
\vec{X}&=\left[\x^{(1)}\ \cdots\ \x^{(N)}\right] \in \Rset^{n \times N}, \\
\label{eq:defy}
\y_f &= \left[y_f^{(1)}\ \cdots\  y_f^{(N)}\right]\T \in \Rset^N,
\end{align}
to denote the input/output data, respectively and
\begin{align}
\label{eq:K}
\K&\!=\!\begin{bmatrix} 	
  k\left(\x^{(1)},\x^{(1)}\right)&\!\cdots\! &k\left(\x^{(1)},\x^{(N)}\right)\\
	\vdots&\!\ddots\!&\vdots \\
   k\left(\x^{(N)},\x^{(1)}\right) &\!\cdots\!&k\left(\x^{(N)},\x^{(N)}\right) 
\end{bmatrix}\! \in\! \Rset^{N \times N}
\end{align}
concatenates kernel evaluations of pairs of input data.
Although the optimization~\eqref{eq:opt_like} is generally non-convex, it is 
usually performed with conjugated gradient-based 
methods~\cite{rasmussen2006gaussian}. Each local minimum can be considered as a 
different interpretation of data and we discuss the effect of suboptimal 
identification in \cref{sec:discussIdentification}.

In a regression task, GPs employ the joint Gaussian distribution of training 
data~$\vec{X},\y_f$ and a test input~$\x^*$
\begin{align}
\label{eq:jointf}
	\begin{bmatrix}  \fGP(\x^*) \\ \y_f \end{bmatrix} \sim \N \left(
			\begin{bmatrix}
			m(\x^*) \\ \mX
			\end{bmatrix},\begin{bmatrix}
				 k^* & \k\T  \\
				\k & \K + \sigon\vec{I}_N  
			\end{bmatrix} \right),
\end{align} 
where
\begin{align}
\label{eq:defmX}
\mX = \begin{bmatrix}
		m\left(\x^{(1)}\right) &\cdots &m\left(\x^{(N)}\right)\end{bmatrix}\T,
\end{align}
to find the posterior mean and variance function 
\begin{align}
\label{eq:mGP}
\mu(\x^*)&:=\E\left[\fGP(\x^*)|\vec{X},\y_f\right] \\
		&\ =m(\x^*) +	 \k\T(\K+\sigon  \nonumber
			 \I_N)^{-1}(\y_f - \mX),  \\
\label{eq:s2GP}
\sigma(\x^*) &:= \V\left[\fGP(\x^*)\right|\vec{X},\y_f]  \\
	&\ = k^* -\k\T(\K+\sigon\I_N)^{-1}\k,\nonumber
\end{align}
through conditioning, where
\begin{align}
\label{eq:defk}
\begin{aligned}
k^* &= k(\x^*,\x^*), \\  
\k&=\left[k\left(\x^{(1)},\x^*\right)\ \cdots \ 
k\left(\x^{(N)},\x^*\right)\right]\T \in \Rset^{N} .
\end{aligned}
\end{align}
However, considering the defined problem in \cref{sec:problem}, the classical 
GP regression framework cannot be directly applied, because closed-loop 
measurements do not provide data points for~$\fdot$ and~$\gdot$ separately. 
Therefore, the following section explains how it is augmented using the given 
prior knowledge.

\subsection{Closed-loop identification with prior knowledge}
\label{sec:GP4ctrlaffinesys}

First, we transfer the knowledge on the positivity of the function~$\gx$ from 
\cref{assum:degree_signg}
into the model~$\ghx$. It is crucial to utilize this knowledge for the model to 
ensure the feedback linearizing control~\eqref{eq:FeLin} results in well 
behaved control signals. Using a GP model for~$\ghx$, this can be ensured using 
a proper prior mean function.
\begin{lem}
	\label{lem:posgh}
	Consider the posterior mean function~\eqref{eq:mGP} with a bounded and 
	differentiable kernel~$k(\cdot,\cdot)$ and a dataset~$(\vec{X},\y_f)$ for 
	which~$\xhiv\neq\x^{(i')}$ and~$\yhi_f>0$, hold~$\forall 
	i,i'=1,\ldots,N,i\neq i'$. Then, there exists a differentiable prior mean 
	function~$m(\x)$ such that 
	\begin{align}
	\mu(\x)>0,\qquad \allx.
	\end{align}
\end{lem}
\begin{IEEEproof}
	Consider a prior mean function for which holds~$0<m\left(\xhiv\right) < 
	\infty$,~$\forall \alli$, then, a differentiable~$m(\x)$ 
	can~$\forall \x \in \Xset\setminus \left\{\x^{(1)},\ldots,\x^{(N)}\right\}$ 
	always be chosen larger than the 
	constant~$\k\T(\K+\sigon\I_N)^{-1}\left(\y_f-\mX\right)$, because
	the latter is bounded. For~$\x\in \left\{\x^{(1)},\ldots,\x^{(N)}\right\}$ 
	a choice~$m\left(\xhiv\right)=\yhi_f$ (which complies with the first 
	condition) ensures, that~$\mu(\x)$ is strictly positive. 
\end{IEEEproof}
\begin{remark}
	\label{remark:posg}
	Since~$\gdot$ is strictly positive by \cref{assum:degree_signg}, the 
	condition~$\yhi_f>0$ 
	follows naturally. In case the Gaussian noise results in negative 
	measurements~$\yhi$, it can 
	be corrected using~$\max(\yhi,\eta)$, with an arbitrarily 
	small~$\eta>0$. 
	Alternatively, strictly positive noise distributions, e.g. a Gamma 
	distribution can also be 
	combined with Gaussian process regression~\cite{rasmussen2006gaussian}. 
	
	In practice, it is 
	often sufficient to set~$m(\x)$ to a 
	positive constant. 
	To verify that~$\mu(\x)>0$ holds, the techniques 
	in~\cite{berkenkamp2016safe} can be utilized. The suitable prior mean 
	function according to \cref{lem:posgh} will be denoted by~$m_g(\x)$.
\end{remark}

Second, the major difficulty of closed-loop identification is to differentiate 
the effect of the control input and the unforced dynamics. For the control 
affine structure, this means that individual measurements of the 
functions~$\fdot$ and~$\gdot$ from~\eqref{eq:sys} are not 
provided. Thus, functions~$\fdot,\gdot$ must be identified from only observing 
their sum exploiting the control affine structure.
We propose to utilize a compound kernels as reviewed in 
Appendix~\ref{sec:kernels} based on~\cite{duvenaud2014thesis}. More 
specifically, we use the 
composite kernel
\begin{align}
	\label{eq:kfg}
	k (\x,\x') = k_{\f}(\x,\x') + u(\x) k_{\g}(\x,\x') u (\x'),
\end{align}
which replicates the structure of a control affine system: the first 
summand~$k_{\f}(\cdot,\cdot)$
represents the unknown unforced dynamics~$\fdot$; the second summand~$u(\cdot) 
k_{\g}(\cdot,\cdot)u(\cdot)$ the 
product of the unknown scaling of the control~$\gdot$ and the known state 
feedback control term~$u(\cdot)$. As no further knowledge 
regarding~$\fdot,\gdot$ is given, we employ two squared exponential~(SE) 
kernels with automatic relevance determination 
\begin{align}
	\label{eq:SEkernelf}
	k_{\f}(\x,\x') &=\sigma_{\f}^2 \exp \left(\sum_{j=1}^n 
	\frac{(x_j-x'_j)^2}{-2l_{j,\f}^2} \right), \\
	\label{eq:SEkernelg}
	k_{\g}(\x,\x') &=\sigma_{\g}^2 \exp \left( \sum_{j=1}^n 
	\frac{(x_j-x'_j)^2}{-2l_{j,\g}^2} \right),
\end{align} 
where the hyperparameters are the lengthscales~\mbox{$l_{j,f}, l_{j,g}\! \in\! 
	\Rset_+$},~$\allj$ and the signal variances~\mbox{$\sigma_{\f},\sigma_{\g} 
	\in 
	\Rset_+^0$}.
For notational convenience, they are concatenated in the vector
\begin{align}
	\bpsi_{gf} = \left[l_{1,f}\ l_{1,g}\ \cdots\  l_{n,f}\ l_{n,g}\ 
	\sigma_f^2\ \sigma_g^2 \right]\T.
\end{align}
The SE kernel is \textit{universal} and therefore allows to model any 
continuous function arbitrarily exactly according 
to~\cite{seeger2008information}.
\begin{remark}
	\label{remark:notparametric}
	GP models with structured kernels, like~\eqref{eq:kfg}, must not be 
	confused 
	with parametric 
	models, which have a predetermined 
	structure and use a fixed number of parameters. In contrast a GP with a 
	structured kernel has potentially infinitely many parameters for each part 
	of its structure. So the kernel encodes the knowledge, that the unknown 
	function e.g. is build of a sum, but each summand has unlimited 
	flexibility. 
\end{remark} 

We denote~$\Uo = \diag\left(u_1\left(\x^{(1)}\right), \dots, 
u_N\left(\x^{(N)} \right)\right) \in \Rset^{N\times N}$, 
where~$u_i$ denotes the control law which was active at the time at which 
the pair~$\left\{\xhiv,\yhi \right\}$ was collected for~$\alli$. 
Furthermore,~$\vec{m}_g^{\vec{X}}$,~$\y$ are analogously defined 
to~\eqref{eq:defmX},~\eqref{eq:defy}, respectively. Then
\begin{align}
\K_{fg} = \K_{f} + \Uo\T\K_{g}\Uo + \sigon\vec{I}_n,
\end{align}
and~$\k_f,\k_g,\K_{f},\K_{g}$ are defined analogously to~\eqref{eq:defk} 
and~\eqref{eq:K} using~$k_f(\x,\x'), k_g(\x,\x')$. 
This notation allows to formulate the estimates~$\fhx,\ghx$.
\begin{lem}
	\label{lem:fhgh}
	The GP posterior mean prediction for the functions~$\fx,\gx$, based on the 
	training data~$\Dk$ in~\eqref{eq:def_Dset} for the compound 
	kernel~\eqref{eq:kfg} are given by
	\begin{align}
	\label{eq:def_fh}
	\fh(\x):=	\mu_f(\x) &= \k_f\T \K_{fg}^{-1}
	\left(\y-\Uo\vec{m}_g^{\vec{X}} \right),\\
	\label{eq:def_gh}
	\gh(\x):=	\mu_g(\x) &=m_g(\x)+ \k_g\T \Uo 
	\K_{fg}^{-1}\left(\y-\Uo\vec{m}_g^{\vec{X}} \right),
	\end{align}
	where the prior mean function for~$\fhx$ is set to zero,~$m_f(\x)=0$, and 
	for~$\ghx$,~$m_g(\x)$ is chosen according to \cref{lem:posgh}. 
\end{lem}
\begin{IEEEproof}
	For an input~$\x$ and the compound kernel~\eqref{eq:kfg}, the joint 
	distribution is given by
	\begin{align}
		\label{eq:jointfg}
		\begin{bmatrix} f(\x) \\ g(\x) \\ \y \end{bmatrix}\!\sim \N \left(
		\begin{bmatrix} 0 \\ m_g(\x) \\ \Uo \vec{m}_g^{\vec{X}} \end{bmatrix},
		\begin{bmatrix} k_f^* & 0 & \k_f\T \\
		0 	  & k_g^* &  \k_g\T\Uo\T  \\
		\k_f & \Uo\k_g  & \K_{fg}
		\end{bmatrix} \right)\!,
	\end{align}
	 similarly to~\eqref{eq:jointf}. According to~\cite{duvenaud2014thesis},
	the posterior mean functions~\eqref{eq:def_fh} and~\eqref{eq:def_gh} follow 
	equivalently 
	to~\eqref{eq:mGP}.
\end{IEEEproof}
For these estimates, it can be shown that all prior knowledge is properly 
transferred into the model.
\begin{prop}
	\label{prop:fhghfollowassum}
	Consider a control affine system~\eqref{eq:sys} under 
	\cref{assum:fgbounded,assum:degree_signg,assum:TrainingData} and the 
	compound kernel~\eqref{eq:kfg}. Then, the estimates~$\fhx$ and~$\ghx$ 
	in~\cref{lem:fhgh} are bounded, infinitely differentiable and there exists 
	a prior mean function~$m_g(\x)$ and a hyperparameter vector~$\bpsi_{gf}$ 
	such that~$\ghx>0$ holds~$\allx$.
\end{prop}
\begin{IEEEproof}
	The SE kernel inherits its properties differentiability and boundedness to 
	all functions represented by the GP~\cite{rasmussen2006gaussian}, thus also 
	to the posterior mean functions, which are used as estimates. The strict 
	positivity 
	of~$\ghx$ follows from the fact, that~$\sigma_g^2$ can be made arbitrarily 
	small such that there always exists a positive function~$m_g$, such that 
	$m_g(\x)$ dominates the term~$\k_g\T \Uo 	
	\K_{fg}^{-1}\left(\y-\Uo\vec{m}_g^{\vec{X}} \right)$ 
	in~\eqref{eq:def_gh}.
\end{IEEEproof}
\begin{remark}
	The only properties of the SE kernel which are used for the derivations and 
	proofs are its differentiability and its boundedness. Thus, the conclusions 
	can directly be extended to other kernel function fulfilling these 
	properties. For the sake of focus, in this article we will	consider the 
	SE kernel only.
\end{remark}

\subsection{Discussion}
\label{sec:discussIdentification}

The most obvious challenge of the closed-loop identification is, that there 
exists not a unique, but infinitely many solutions for two differentiable 
functions to add up to the same values. Thus, only observing the sum 
in~\eqref{eq:sys} is not promising to learn the unique correct individual 
functions~$\fdot$,~$\gdot$ because it is an under-determined problem. The 
estimates in~\eqref{eq:def_fh} and~\eqref{eq:def_gh} are just one of many 
solutions, determined by 
the choice of hyperparameters, which suits the training data. Nevertheless, the 
optimization~\eqref{eq:opt_like} interprets the observed data to match the 
kernel structure, which is shown to be successful in the simulation in 
\cref{sec:simSec1}. For the case that the results are not satisfactory, we 
provide an extension in Appendix~\ref{sec:improveid} to address this challenge. 
It merges data points of the closed-loop system with measurements from the 
temporary open-loop system. It thereby uses \cref{lem:finiteEscapetime}, 
which allows to safely turn off the control signal ($u=0$) for a finite 
time period. Nevertheless, we want to highlight, that the formal guarantees 
provided in the following section (\cref{thm:UUB}) hold independently whether  
this extension is utilized or not.

Furthermore, additional knowledge like periodicity or dependence of~$\fdot$ 
or~$\gdot$ on only a subset of the state variables, can also be transferred 
into the kernel to facilitate the identification by using a periodic kernel or 
setting the lengthscales~$l_{j,f}= l_{j',g}=\infty$ for the states~$j,j'$ of 
which they are independent, respectively. The latter simplifies the 
optimization of hyperparameters as the search space is reduced. A systematic 
way of constructing more evolved kernel function (including more prior 
knowledge) is discussed in~\cite{duvenaud2014thesis}. 

Considering the computational load, the inverse of~$\K_{fg}$ is most critical, 
as the number of operations increases cubical with the number of data points, 
thus~$\O(N^3)$. However, adding further data points is necessary to ensure the 
model is precise at the current position in the state space, where the most 
recent data points are taken from measurements. Comparing to previous 
approaches, e.g.~\cite{umlauft2017feedback}, where~$\K_{fg}^{-1}\y$ is constant 
and can thereby be precomputed offline, here it must be recomputed with every 
update of the model as data points are added one at a time. This difficulty can 
be addressed using a \mbox{rank-$1$} update of the inverse with the 
Sherman–Morrison formula~\cite{sherman1950adjustment} resulting in only 
~$\O(N^2)$ operations. However, this quadratic computational complexity might 
still be very time consuming, and therefore motivates the data-efficient 
event-triggered model updates introduced in \cref{sec:event}.

Generally, Gaussian processes turn out to be very effective for the adaptive 
model control law: They properly transfer all prior assumptions consistently 
into the model (\cref{prop:fhghfollowassum}) and allow for the identification 
in closed-loop. The nonparametric nature allows an unlimited model flexibility 
and the complexity increases as more data is 
available in a data-driven fashion. This is a crucial advantage compared to 
classical system 
identification methods, particularly for highly nonlinear systems.

\section{Feedback Linearizing Control Law}
\label{sec:control}
In this section, the feedback linearizing online learning control law is 
proposed and the resulting closed-loop behavior is analyzed. After showing 
ultimate boundedness for the most general case, we make further specific 
assumptions to provide stronger stability results. Here, further properties of 
the Gaussian process modeling technique is exploited: As the model error of the 
GP can be bounded and quantified, the ultimate bound of the tracking error can 
also be quantified.

Classical model reference adaptive control modifies the model parameters 
continuously over time, which is not possible here due to the nonparametric 
nature of the GP model. Thus, particular attention must be drawn to the 
resulting 
switching character of the control law, which stems from the time-varying 
dataset~$\Dk$ of the Gaussian process introduced in 
\cref{assum:TrainingData}.

We are interested in tracking desired trajectories for the state~$x_1$, given 
by~$x_d(t)$, with the following  property.
\begin{assum}
	\label{assum:x_d}
	The desired trajectory~$x_d(t)$ is bounded and at least~$n-1$ times 
	differentiable, thus 
	\begin{align}
		\x_d(t) = \begin{bmatrix}x_d &\dot{x}_d &\cdots& 
		\frac{d^{n-1}x_d}{dt^{n-1}}	\end{bmatrix}\T
	\end{align} 
	is continuous and~$\frac{d^{n}x_d}{dt^{n}}$ is bounded.\footnote{The~$t$ 
	dependency of ~$\x_d,\x$ and the~$\x$ dependencies of~$u,\f,\fh,\g,\gh$ are 
	partially omitted for notational 
		convenience.}
\end{assum}
For notational convenience, we define the tracking error
\begin{align}
	\e = \x - \x_d.
\end{align}

\subsection{Control law}
\label{sec:ctrl_law}
\begin{figure*}
	\vspace{1.5mm}
	\tikzstyle{block}=[draw, inner sep=2, outer sep=0, minimum width=2em, 
	rounded corners, thick,  align=center]
	\centering  \tikzsetnextfilename{ctrl_struct_eventonline}
	\tikzset{external/export next=false}
	\begin{tikzpicture}[node distance=2cm, 		
		auto,>=latex',every text node 
		part/.style={align=center},thick,scale=1, every 
		node/.style={scale=1}]
		\def\dis{0.6}
		\def\sh{0.1}
		\node [block] (sys) {control affine \\ system~\eqref{eq:sys} };
		\node [block,left = \dis of sys] (FeLin) {feedback \\ linearization 
		\eqref{eq:FeLin} };
		\node [block,below =\dis  of sys] (fgh) {GP model \\ 
		\eqref{eq:def_fh}~\eqref{eq:def_gh}};	
		\node [block,right=\dis  of fgh] (event) {model update\\event trigger 
		\eqref{eq:event}};	
		\node [circ, left = \dis  of FeLin] (sum) {};  	
		\node [block,left = \dis  of sum] (kc) {$\kc$};	
		\node [block,left = \dis  of kc] (lambda1) {$[\ckb\T  \ 	1]$};		
		\node [block,below =\dis  of kc] (lambda0) {$[0\ \ckb\T ]$};
		\node [circ, left = \dis  of lambda1] (sume) {};  	
		\node [block,below =\dis  of fgh] (data) {$\Dk$};
		
		\draw[<-] (sume.west) --++ (-\dis,0) 
				node[pos=1,anchor=east] (dxn){$\x_d$}
				node[pos=0.4,anchor=south] {$-$};
		\draw[->] (sume.east)--(lambda1.west)node[pos=0.5,anchor=south] 
		(e){$\e$};
		\draw[->] (lambda1)--(kc) node[pos=0.5,anchor=south] (lambda12kc){$r$};
		\draw[->] (kc)--(sum) node[pos=0.5,anchor=south] (kc2sum){$-$};
		\draw[->](lambda0)-|(sum) node[pos=0.3,anchor=south]{$\yd$}
								  node[pos=0.9,anchor=east]{$-$};
		\draw[->] (sum)--(FeLin) node[pos=0.5,anchor=south] (sum2FeLin){$\nu$};
		\draw[->] (FeLin)--(sys) node[pos=0.5,anchor=south] (Felin2sys){$u$};
		\draw[->] ([shift ={(0,\sh)}]fgh.west)-|
				  ([shift ={(\sh,0)}]FeLin.south) 
					node[pos=0.3,anchor=south] 	(fh){$\fhk$};
		\draw[->] ([shift ={(0,-\sh)}]fgh.west)-|
				  ([shift ={(-\sh,0)}]FeLin.south) 
				node[pos=0.5,anchor=east] 	(gh){$\ghk$};
		\draw[->] (data)--(fgh) node[pos=0.5,anchor=south] (data2fgh){};
		\draw[->] (sys.east)-|++ ($(fgh)+(0.3,\sh)$) node[pos=0.5,anchor=west] 	
		(x1){$\x$} --	([shift = {(0,\sh)}]fgh.east) 	  ;
		\draw[->] ([shift={(0,-\sh)}]fgh.east)--([shift ={(0,-\sh)}]event.west)
				node[pos=0.5,anchor=north] 	(x1){$\sigma_\kap$} ;
		\draw[<-o] ([shift = {(0,\sh)}]data.east)--++ (2.2,0);
		\draw ([shift = {(2.3,\sh+0.3)}]data.east) --++ (0.3,-0.3)--++ (1,0)
				node[pos=1,anchor=west] (x){$\x$};
		\draw[<-o] ([shift = {(0,-\sh)}]data.east) --++ (2.2,0);
		\draw ([shift = {(2.3,-\sh+0.3)}]data.east) --++ (0.3,-0.3)--++ (1,0)
				node[pos=1,anchor=west,shift={(0,-\sh)}] (dxn){$\dot{x}_n$};
		\draw[->] (event.south) --++ (0,-0.25);		
		\draw ([shift = {(1.8,-0.3)}]data.east) rectangle 
			  ([shift = {(2.9,0.5)}]data.east);  
		\draw[->] (sys.east)-| ([shift = {(0.3,1.1cm)}]sys.east) -| (sume.north)
			node[pos=0.5,anchor=south] 	(x2){$\x$};
		\draw[->] ([shift = {(-0.3,0)}]lambda1.west) |- (lambda0.west)	  	
			node[pos=0.9,anchor=south] 	(x3){$\e$};			
		
		\draw[red, dashed] ([shift={(-0.1,0.1cm)}]FeLin.north west) 
			rectangle ([shift={(0.15cm,-0.1cm)}]sys.south east);
		\draw[blue, dashed] ([shift = {(-0.5cm,0.1cm)}]FeLin.north west)
			rectangle	([shift = {(-2cm,-0.1cm)}]lambda0.south west);
		\node[anchor = west, red, align = left] at 
		([shift={(0,0.4)}]FeLin.north west) 
			{linearized for	$\fhk= f,\ \ghk=g$};
		\node[anchor=north,blue,align=left] at ([shift={(0,-0.2)}]lambda0.south)
			{linear control law \\$\ckb$ is Hurwitz};		
	\end{tikzpicture}
	\caption{The online leaning feedback linearizing control scheme including 
			the event trigger proposed in \cref{sec:event}, which controls the 
			switching time~$\tkp$.}
	\label{fig:ctrl_struct_eventonline}
\end{figure*}
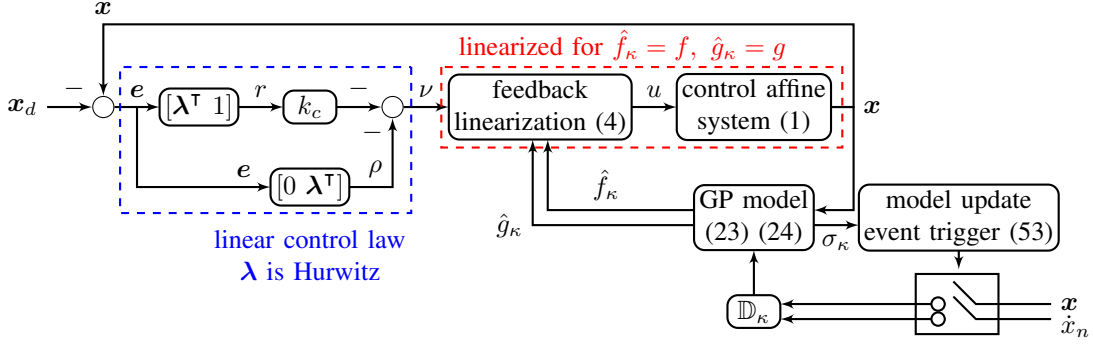

Consider the filtered scalar state~$r \in \Rset$, defined as
\begin{align}
	r = \begin{bmatrix} \ckb\T &1 \end{bmatrix}\e,
\end{align}
where~$\ckb = [\ck_1\ \ck_2 \cdots \ck_{n-1}]\T \in \Rset^{n-1}$ is a 
coefficient 
vector such that for~$s \in \mathbb{C}$ the polynomial \mbox{$s^{n-1} + 
\lambda_{n-1} s^{n-2} + \cdots + \lambda_1$} is Hurwitz. Under this condition, 
the error converges exponentially~$\e \to \vec{0}$ as~$r\to 
0$~\cite{yecsildirek1995feedback}. The dynamics of the filtered state is
\begin{align}
	\dot{r} = \fx + \gx u(\x) +\yd,
\end{align}
where
\begin{align}
	\yd = \ckb\T \e_{2:n}-\frac{d^{n}x_d}{dt^{n}} ,
\end{align}
with~$\e_{2:n} = \left[e_2 \cdots e_n\right]\T \in \Rset^{n-1}$. 
For the control law~$u(\x)$, we propose
\begin{align}
\label{eq:uf}
u_\kap (\x) = \frac{1}{\ghkx}\left(-\fhkx -\kc r  -\yd \right),
\end{align}
according to~\eqref{eq:FeLin} where~$ \nu = -\kc r  -\yd$
with~$\kc \in \Rset_+$ is used. The subscript~$\kap\in \Nset_0$ indicates 
the~\mbox{$\kap$-th} time interval~\mbox{$t\in [\tk \  \tkp)$} for 
which~$u_\kap$ is applied according to~\eqref{eq:uswitch}. The 
estimates~$\ghkdot$,~$\fhkdot$ are based on~$N_\kap$ training points in the 
time-varying dataset~$\Dk$ introduced in \cref{assum:TrainingData}. The 
control scheme is visualized in \cref{fig:ctrl_struct_eventonline} and the 
adaptation procedure is provided in \cref{alg:adaptive_ctrl}.

Note, that even though GPs itself are probabilistic models, the control law is 
deterministic, because it only employs the posterior mean functions as model 
estimate~$\ghkdot$,~$\fhkdot$.

\begin{algorithm}
	\begin{algorithmic}[1]
		\State initialize~$\kap=0$,~$\Dset_0 = \{ \}$,~$\fh_0= 0$,~$\gh_0= 
		m_g(\cdot)$
		\While {simulation time not exceeded}
			\While{$t<\tkp$}
				\State run controller~$u_\kap$ in~\eqref{eq:uf}
			\EndWhile
		\State set~$\kap\leftarrow \kap+1$
		\State measure~$\x^{(\kap)} = \x(\tk)$ and~$y^{(\kap)} = 
		\dot{x}_n(\tk) +\epsilon^{(\kap)}$ 
		\State add training point~$\Dk = \Dset_{\kap-1} \cup 
		\left\{\left(\x^{(\kap)},y^{(\kap)}\right)\right\}$
		\State update the estimates~$\fhkdot,\ghkdot$ 
		in~\eqref{eq:def_fh},~\eqref{eq:def_gh}
		\EndWhile
	\end{algorithmic}
\caption{Online leaning for feedback linearization control}
\label{alg:adaptive_ctrl}
\end{algorithm}

\subsection{Convergence analysis}
\label{sec:stabana}
An offline version of the control law~\eqref{eq:uf}, with constant 
dataset~$\Dset$ and estimates~$\fhx,\ghx$ was introduced 
previously and was shown to be globally uniformly ultimately 
bounded~\cite[Proposition 1]{umlauft2017feedback}.

But, with the time-varying dataset and model, \cref{alg:adaptive_ctrl} 
describes a switching control law. This switching results in a hybrid system, 
where some states are changing continuously in time but the system dynamics (or 
other states) change at discrete time instances. Here, the resulting 
closed-loop system is subject to (in general) arbitrary switching. Its 
convergence behavior is analyzed here based on the principle of a common 
Lyapunov function. It states, that a Lyapunov function, which is independent of 
the switching signal must decrease over time along the system's trajectories. 
This is shown in the following.
\begin{thm}
	\label{thm:UUB}
	Consider the system~\eqref{eq:sys} and a desired trajectory~$\x_d(t)$ under 
	\cref{assum:fgbounded,assum:TrainingData,assum:degree_signg,assum:x_d}.
	Further consider the control law~\eqref{eq:uf}, where~$\fdot,\gdot$ are 
	modeled by GP mean functions~$\fhkdot,\ghkdot$ in~\eqref{eq:def_fh} 
	and~~\eqref{eq:def_gh}, respectively. The GP model is updated at
	arbitrary switching times~$\tk$ according to \cref{alg:adaptive_ctrl}.  
	Then, there exists a~$\kc^*>0$ such that for every~$\kc \geq \kc^*$ the 
	tracking error~$\norm*{\e}$ is globally uniformly ultimately bounded.
\end{thm}
\begin{IEEEproof}
	Consider the common Lyapunov function candidate
	\begin{align}
		\label{eq:V}
		V_\kap(\x) = r^2/2, \qquad \forall \kap\in \Nset_0,
	\end{align}
	with time derivative
	\begin{align}
	\label{eq:dotV}
	\dot{V}_\kap(\x) 
	&= r\dot{r}= r \left(\f + \g u_\kap  +\rho \right)  \\
	&= r \left(\f +\frac{\g}{\ghk}(-\fhk -\kc r - \yd) 
	+ \yd \right)  \nonumber \\
	&= r \left(\f - \gb_\kap\fhk\right)- \kc \gb_\kap r^2+ (1-\gb_\kap) r 
	\yd,
	\nonumber	
	\end{align}
	where~$\gb_\kap:= \frac{\gx}{\ghk(\x)}$ is positive and bounded~$\forall 
	 \kap$ and~$\allx$ from \cref{prop:fhghfollowassum} and 
	\cref{assum:fgbounded,assum:degree_signg}. As a 
	consequence~\mbox{$\left(\f - \gb_\kap\fhk\right)$} is bounded and 
	there 
	exists a constant~\mbox{$\vec{a} \in \Rset^n$} such that 
	\begin{align}
		\norm*{r \left(\f - \gb_\kap\fhk\right)} \leq \norm*{\vec{a}\T\e} 
		\qquad 
		\forall \e,\kap
	\end{align}
	holds, because~$r$ only grows linearly in~$\e$. For similar reasons, we can 
	find constants~$\vec{c} \in \Rset^n$ and~$\vec{B},\vec{C} \in 
	\Rset^{n\times n}$, with~$\vec{B},\vec{C} \succ 0$ for which   
	\begin{align}
	\label{eq:BCcbounds}
	 \norm*{\gb_\kap r^2} &\geq \e\T \vec{B} \e,  	 &\forall \e,\kap \\
	 \norm*{(1-\gb_\kap) r \yd} &\leq \e\T \vec{C} \e + \vec{c}\T \e,   
	 &\forall \e,\kap
	\end{align}
	holds which exist since~$\f, \fhk, \gb_\kap$ are bounded. Therefore, 
	\begin{align}
		\dot{V}_\kap(\x) &\leq\! \norm*{\vec{a}} \norm*{\e} 
				\!-\! \kc \smin(\vec{B}) \norm*{\e}^2 
				\!+\! \smax(\vec{C}) \norm*{\e}^2\nonumber
				\!+\! \norm*{\vec{c}} \norm*{\e} \\
				& =\! \norm*{\e} 
				\!\left(\norm*{\vec{a}}\!+\!\norm*{\vec{c}}\right) 
				\!+\! \norm*{\e}^2\!\left(\smax(\vec{C}) \!-\!\kc 
				\smin(\vec{B})\right)
	\end{align}
 	holds for all~$\kap$ and there exists a~$\kc^*>0$  such that 
	 \begin{align}
	 	\label{eq:condAB}
		\smax(\vec{C})- \kc^* \smin(\vec{B})<0. 
	\end{align}
	As a result, for every~$\kc\geq\kc^*$,	the Lyapunov function decreases
	\begin{align}
	\dot{V}_\kap(\x)<0, \qquad \allx\!\setminus\!\Bset,\ \forall \kap
	\end{align} 
	outside of the set
	\begin{align}
	\label{eq:def_Bset}
		\Bset = \left\{\allx \left| \norm*{\e} \leq 
		\frac{\norm*{\vec{a}}+\norm*{\vec{c}}}{\kc \smin(\vec{A}) 
		-\smax(\vec{B})}\right.\right\},
	\end{align}
	which forms a tube in $\x$ coordinates around the desired trajectory and a 
	ball in~$\e$ coordinates.
	Thus, we have found a common radially unbounded Lyapunov function~$V(\x)$, 
	which decreases~$\forall \kap$ outside of the ball~$\Bset$. According 
	to~\mbox{\cite[Theorem 2.1]{liberzon2012switching}}, this allows the 
	conclusion 
	that for arbitrary switching sequences the tracking error converges 
	 to the ball~$\Bset$. Since~$\Bset$ is independent of the initial 
	 state, global uniform ultimate boundedness (GUUB) holds.
\end{IEEEproof}
Thus, we have shown, that the tracking error is bounded by the proposed 
online learning control scheme for a large enough gain~$\kc$ 
with an arbitrary 
switching sequence. However, without further 
knowledge, a value for the critical gain~$\kc^*$ cannot be computed.
We therefore make the following simplifying assumption
\begin{assum}
	\label{assum:perfectg}
	The function~$\gx$ is known, thus~\mbox{$\ghx= \gx$} and 
	noisy training data observations of~$\dot{x}_n$
	\begin{align}
	y_{f}^{(i)}  = \fxi + \epsilon^{(i)} 
	=\dot{x}_n^{(i)} - \gxi u_\kap \left(\xhiv \right) + \epsilon^{(i)} 
	\end{align}
	with~$\epsilon^{(i)} \sim \N(0,\sigon),\ i=1,\ldots,N_\kap$ are available.
\end{assum}
This assumption holds for many real-world system, e.g. most Lagrangian systems, 
which is a considerably large class~\cite{beckers2019stable}. It is also a 
quite common assumption when 
working 
with control affine systems~\cite{slotine1993robust}. 

Analogously to~\eqref{eq:def_fh}, the unknown function is now estimated by
\begin{align}
\label{eq:fh}
\fhx:= \mu_f(\x) =  \k_f\T(\K_f+\sigon \I_{N_{\kap}})^{-1}\y_f,
\end{align}
where~$\y_f = \left[y_{f}^{(1)} \cdots y_{f}^{(N_\kap)} \right]\T$ and~$\k_f$, 
$\K_f$ are computed according to~\eqref{eq:defk} and~\eqref{eq:K} for the 
SE kernel. 
Under this assumption, \cref{thm:UUB} can be relaxed with respect to the 
choice of the gain~$\kc$.
\begin{cor}
	\label{cor:UUB_perfectg}
	Consider the system~\eqref{eq:sys} and a desired trajectory~$\x_d(t)$ under 
	\cref{assum:fgbounded,assum:TrainingData,assum:degree_signg,assum:x_d,assum:perfectg}.
	Further consider the control law~\eqref{eq:uf}, where~$\fdot$ is modeled by 
	a GP mean function~$\fhkdot$ in~\eqref{eq:fh}, which is adapted at 
	arbitrary switching times~$\tk$ according to \cref{alg:adaptive_ctrl}. 
	Then, the tracking error~$\norm*{\e}$ of the closed-loop switching system 
	is globally uniformly ultimately bounded for any~$\kc > 0$.
\end{cor}
\begin{IEEEproof}
	Using the Lyapunov function~\eqref{eq:V}, we obtain for~$\gb_\kap=1$
	\begin{align}
	\label{eq:dotVnog}
		\dot{V}_\kap(\x) 
		&= r\dot{r}= r \left(\f + \g u_\kap  +\rho \right)  \\
		\label{eq:dotVnog2}
		&= r \left(\f - \fhk\right)- \kc  r^2,
	\end{align}	
	which leaves us with the condition~$\left(\f - \fhk\right)< \kc  r$ for 
	negative definiteness of~$\dot{V}_\kap(\x)$. Thus independent of the 
	gain~$\kc$, their exists a 
	ball outside of which the Lyapunov function is decreasing~$\forall \kap$, 
	which leads to the GUUB~$\forall \kc >0$.
\end{IEEEproof}

For completeness, we also formalize the result for dataset~$\Dk$ which 
remains constant after~$\tk$, thus no further measurements are taken and no 
data points are deleted from the set for~$t>\tk$.
\begin{cor}
	\label{cor:UUB_perfectg_constantD}
	Consider the system~\eqref{eq:sys} and a desired trajectory~$\x_d(t)$ under 
	\cref{assum:fgbounded,assum:TrainingData,assum:degree_signg,assum:x_d,assum:perfectg}.
	Further consider the control law~\eqref{eq:uf}, where~$\fdot$ is modeled by 
	a GP mean function~$\fhkdot$ in~\eqref{eq:fh} with a fixed dataset~$\Dk$.
	Then, the tracking error~$\norm*{\e}$ of the closed-loop system 
	is globally uniformly ultimately bounded for any~$\kc > 0$.
\end{cor}
\begin{IEEEproof}
	The proof is straightforward as it follows from the proof for 
	\cref{cor:UUB_perfectg}.
\end{IEEEproof}

\subsection{Quantifying the ultimate bound}
\label{sec:Prox_mtotrue}
\Cref{thm:UUB,cor:UUB_perfectg,cor:UUB_perfectg_constantD} show, that there 
exists an ultimate bound for the tracking error~$\e$, however, its size is 
unknown. To quantify the ultimate bound~$\Bset$, an upper bound for the model 
estimate, defined as
\begin{align}
\label{eq:def_df}
\df_\kap(\x)=|\fx - \fhkx|,\quad \forall \kap,
\end{align}  
is derived in this section using the variance function~$\sigma_\kap(\x)\colon 
\Xset \to \Rset_+^0$ of the GP as defined in~\eqref{eq:s2GP}.
Since the GP is a probabilistic model in nature, we cannot expect any 
deterministic statements regarding the error of the estimate~$\df_\kap$.
However, according to~\cite{srinivas2012information}, it is possible to make 
high probability statements regarding the maximum distance from the true 
function~$\fx$ to the mean function~$\mu(\x)$ on a compact set. As
known from the no-free lunch theorems~\cite{wolpert2002supervised}, this 
generalization cannot be expected without any prior knowledge about~$\fdot$. 
Since we do not want to make any parametric assumptions which limit the 
complexity of~$\fdot$, we restrict its reproducing kernel Hilbert space (RKHS) 
norm as follows. 
\begin{assum}
	\label{assum:RKHSnorm}
	The function~$\fx$ has a bounded reproducing kernel Hilbert space (RKHS) 
	norm with respect to a squared exponential kernel~$k(\cdot,\cdot)$, 
	with known hyperparameters
	denoted by~\mbox{$\norm{\fx}_k^2 \leq B_f$}. 
\end{assum}
With this additional assumption, a high probability statement regarding 
precision of the mean function estimate is possible according 
to~\cite{srinivas2012information}. 
\begin{lem}	\label{lem:mtotrue}
	Suppose \cref{assum:RKHSnorm} holds, then 
	\begin{align}
	\label{eq:mtotrue}
	\Pr \left\{\left| \mu_\kap (\x) \!- \!\f (\x) \right|
		 \leq 	\beta_\kap  \sigma_\kap(\x),\allxt\!,\!N_\kap\in\Nset_0 
		 \right\} 
		 \geq 
		 1\! -\! \delta,
	\end{align}
	holds on a compact set~$\Xsett \subset \Rset^n$, where~$\delta \in (0,1)$, 
	$\beta_\kap = \sqrt{2B_f + 300\gamma_\kap \log^3((\kap+1)/\delta)}$ and 
	$\gamma_\kap$ is the maximum mutual information that can be obtained 
	about~$\fdot$ from~$\kap+1$ noisy samples~$\x_{(1)}, \ldots, \x_{(\kap+1)}$ 
	and~$\mu_\kap(\x)$ and~$\sigma_\kap(\x)$ are posterior mean and variance 
	function of a GP for~$N_\kap$ data points as defined in~\eqref{eq:mGP} and
	\eqref{eq:s2GP}, respectively. 
\end{lem}
\begin{IEEEproof}
	This is a direct consequence from~\cite[Theorem~6]{srinivas2012information}.
\end{IEEEproof}
\begin{remark}
	\label{remark:not1}
	Consider, that~\eqref{eq:mtotrue} takes all~$N_\kap\in\Nset_0$ into account 
	at once. This means, the probability~$\delta$ holds not 
	just for a single~$N_\kap\in\Nset_0$ but for all~$N_\kap\in\Nset_0$. This 
	becomes clear when rewriting~\eqref{eq:mtotrue} as
	\begin{align}
	\Pr \left\{\bigcap_{N_\kap=0}^\infty \left| \mu_\kap (\x) - \f (\x) \right|
	\leq \beta_\kap  \sigma_\kap(\x),\allxt\right\} \geq 1 - \delta.
	\end{align}
\end{remark}

The model error bound in \cref{lem:mtotrue} only holds on a compact 
set~$\Xsett$. Nevertheless, we have already shown in 
\cref{thm:UUB,cor:UUB_perfectg,cor:UUB_perfectg}, that the tracking error 
converges to a compact 
set~$\Bset$. Thus, we set~$\Xsett=\Bset$, which leads to the following result.
\begin{thm}
	\label{thm:compB}
	Consider the system~\eqref{eq:sys} and a desired trajectory~$\x_d(t)$ under 
	\cref{assum:fgbounded,assum:TrainingData,assum:degree_signg,assum:x_d,assum:perfectg,assum:RKHSnorm}.
	Further consider the control law~\eqref{eq:uf}, where~$\fdot$ is modeled by 
	a GP mean function~$\fhkdot$ in~\eqref{eq:fh}, which is adapted at 
	arbitrary switching times~$\tk$ according to \cref{alg:adaptive_ctrl}.
	Then, with probability~\mbox{$1-\delta$},~$\delta\in (0,1)$, the tracking 
	error~$\norm*{\e}$ is uniformly ultimately bounded for any~$\kc > 0$ with 
	the ultimate bound 
	\begin{align}
	\Bset_\kap = \left\{\allx\left| \norm{\e} 
	\leq\frac{\beta_\kap \bar{\sigma}_\kap}{\kc\left\| \begin{bmatrix} \ckb\T & 
	1 
		\end{bmatrix}\right\|}\right.\right\}, \quad \forall \x_0 \in \Xset,
	\end{align}
	where~\mbox{$\bar{\sigma}_\kap\coloneqq \max_{\x \in \Xsett} 
	\sigma_\kap(\x)$} 
	and~$\beta_\kap$ is defined in \cref{lem:mtotrue}.
\end{thm} 
\begin{IEEEproof}
	Using the common Lyapunov candidate~\eqref{eq:V}, its time 
	derivative~\eqref{eq:dotVnog2} is given in the~$\kap$-th time step for the 
	case~$\g-\gh=0$ (\cref{assum:perfectg}) by
	\begin{align}
	\label{eq:dotVgb1}
	\dot{V}_\kap(\x) \leq  r \df_\kap(\x) -  \kc r^2.
	\end{align}
	As \cref{thm:UUB} guarantees convergence to~$\Bset=\Xsett$, the model error 
	must only be closer considered in this compact set. From \cref{lem:mtotrue} 
	it can be concluded, that 
	\begin{align}
		\Pr\left\{\df_\kap(\x) \leq \beta_\kap \bar{\sigma}_\kap, 
		\allxt, \kap\in\Nset_0 \right\} \geq  1-\delta  \\
		\Rightarrow \Pr\left\{\dot{V}(\x) < 0,
		\allxt \!\setminus\! \Bset_\kap, \kap\in\Nset_0 \right\} \geq  
		1-\delta ,
	\end{align}
	which shows convergence of~$r$ to a ball with radius~$\frac{\beta_\kap 
	\bar{\sigma}_\kap}{\kc}$ and the error is~$\e$ is ultimately bounded 
	by~$\Bset_\kap$ with probability larger then~$1-\delta$. The attributes 
	hold uniformly and globally from the fact that~$V_\kap$ is a common, 
	time-independent and radially unbounded Lyapunov 
	function~\cite{liberzon2012switching}.
\end{IEEEproof}

\begin{remark}
	\label{remark:UUBvscompB}
	\Cref{thm:UUB} focuses on the existence of an ultimate bound and therefore 
	provides with~$\Bset$ in~\eqref{eq:def_Bset} the maximum bound across all 
	time intervals, which can be seen in~\eqref{eq:BCcbounds}, 
	where~$\vec{B},\vec{C},\vec{c}$ must be suitable for all~$\kap$.
	In contrast, \cref{thm:compB} is here more specific and provides 
	with~$\Bset_\kap$ a quantitative bound for each time interval~$\kap$ 
	individually. Note, that the tracking error~$\e$ will not necessarily 
	converge to the ball~$\Bset_\kap$ by the end of the~$\kap$-th time 
	step~$\tkp$ because it might take 
	infinite time. 	It is considered as a ball which is reached by the 
	tracking error if the control law stops adapting after the~$\kap$-th update 
	(compare \cref{cor:UUB_perfectg_constantD}).
\end{remark}

\begin{remark}
	In contrast to \cref{thm:UUB}, \cref{thm:compB} is a stability statement 
	which only holds with a specified probability. The reason for this lies in 
	the uncertainty about the plant itself, but neither the plant nor any part 
	of the controller are stochastic. Therefore, a stability analysis from 
	deterministic control theory is applicable here, but the convergence result 
	does not hold for all plants which fulfill the specified assumptions. There 
	exists a small fraction of all plants (specified by~$\delta$), which do not 
	converge to the specified ultimate bound. But, if the plant does not belong 
	to this fraction, the result will always hold and there is no stochastic 
	stability analysis necessary.
	Note, that the fraction~$\delta$ for which the result does not hold, can be 
	be made arbitrarily small.
\end{remark}

\section{Event-triggered Model Update}
\label{sec:event}
The results in the previous section all hold for arbitrary switching 
sequences~(any definition of~$\tkp$ is possible)
because we have so far not specified when new training data 
points are 
taken and the model is updated accordingly. Our goal is a data-efficient online 
learning scheme and thereby we only want to add training data if necessary. 
Thus, 
switching should not 
occur synchronously (after a specified fixed time interval) but asynchronously 
(whenever needed), which is known as event-triggered control.

The general idea of event-based control is to utilize a scarce resource (sensor 
measurements, computational power, communication channel, etc.) only when 
required. In contrast to time-triggered control, where the resource is used 
periodically (synchronously), it is thereby typically more 
resource-conserving~\cite{heemels2012introduction}. In our setting, we aim to 
reduce the number of the model updates and measurements for training data to 
keep the computational complexity low. The key idea of our data-efficient 
online learning is therefore, to take only new training data into account if 
there is necessity based on the current uncertainty in the model.

Previous work in~\cite{chowdhary2015bayesian} uses a 
time-triggered model, thus measurements are taken and training points are added 
after a specific time, thus~$\tkp = \tk + \dt$ with fixed interval~$\dt>0$. 
However, this causes the following difficulties: First, it is unknown whether 
the current estimate~$\fhkdot$ of the function~$\fdot$ is precise enough to 
ensure a further decrease of the Lyapunov function. From~\eqref{eq:dotVgb1} it 
is clear that, the estimates must become more precise as~$r$ gets smaller to 
guarantee negative definiteness of $\dot{V}_\kap$ for~$\allx$. 

Considering, that at some parts of the state space more data points are 
necessary to model the function~$\fdot$ precisely than in others shows that 
choosing a constant~$\dt$ properly is impossible without knowing the 
function~$\fdot$.
Second, over an infinite time horizon, the time-triggered update causes 
infinitely many (possibly unnecessary) measurements. 
This is critical even for finite time, because the number of operations to 
update the GP model increases with~$\O(N^3)$ (or~$\O(N^2)$ at 
best, when using the Sherman–Morrison formula)~\cite{rasmussen2006gaussian}.  

In summary, for the time-triggered design, there is a trade-off between the 
precision and the computational complexity of the model when choosing the 
update rate. If more points are added to the dataset, the variance of the GP 
model and thereby the maximum model error decreases according to 
\cref{lem:mtotrue}. However, many training points increase the time to compute 
the model estimate and possible result in a loss of the real-time 
capability~\cite{nguyen2009local}. 

Therefore, in interest of data efficiency and the associated computational 
complexity, we trigger measurements and their intake to the dataset in an
event-based fashion. An intuitive idea is to add training points as soon as the 
error~$~\df_\kap$ becomes too large, which is specified based on 
the Lyapunov stability condition. Generally, to guarantee stability, an event 
must triggered before the temporal derivative of the Lyapunov function turns 
non-negative, thus 
\begin{align}
\label{eq:eventGeneral}
\tkp \coloneqq \left\{t>\tk\left| 
\dot{V}(\x)\geq 0\right. \right\}.
\end{align}
However, since an exact computation of~$\dot{V}(\cdot)$ is not possible, we 
have to evaluate an upper bound as presented in the following.
First, we will consider noiseless measurements of the highest state 
derivative for the training data, 
before addressing the case, where these measurements are corrupted by noise.

\subsection{Asymptotic stability for noiseless measurements}
\label{sec:noiseless}

We first define the noise free case formally in an assumption.
\begin{assum}
	\label{assum:noisefree}
	Measurements of~$\dot{x}_n$ are available noise free, thus~$\sigon=0$ in 
	\cref{assum:TrainingData}.
\end{assum}
A well-suited indicator for the necessity to add a new training point is the 
variance function of the GP~$\sigma_\kap(\cdot)$ in~\eqref{eq:s2GP} as it 
bounds the maximum error with high probability according to \cref{lem:mtotrue}. 
Based on this intuition, we propose the following event
\begin{align}
\label{eq:event}
	\tkp \coloneqq \left\{t>\tk\left| 
	\beta_\kap\sigma_\kap(\x)\geq\kc |r|\right. \right\},
\end{align}
where the triggering time~$\tkp$ is defined as the first time 
after~$\tk$ when~$\beta_\kap\sigma_\kap(\x)$ becomes 
larger or equal than~$\kc|r|$.
\begin{remark}
	In the time instance after each update~$t=\tk$, it generally 
	holds~$\sigma_{\kap}(\x(\tk))=0$, which 
	implies~$\sigma_{\kap}(\x(\tk))\leq\kc |r(\tk)|$. 
	Since both~$\sigma_\kap$ and~$r$ are continuous over time between two 
	events, the event will always be triggered at the equality, 	
	thus~\mbox{$\beta_\kap\sigma_\kap(\x(\tkp))=\kc |r(\tkp)|$}.
\end{remark}
Using the proposed event~\eqref{eq:event} in \cref{alg:adaptive_ctrl} as 
trigger to update the model, the following is concluded.
\begin{thm}
	\label{thm:asymEvent}
	Consider the system~\eqref{eq:sys} and a desired trajectory~$\x_d(t)$ under 
	\cref{assum:fgbounded,assum:TrainingData,assum:degree_signg,assum:x_d,assum:perfectg,assum:RKHSnorm,assum:noisefree}.
	Further consider the control law~\eqref{eq:uf}, where~$\fdot$ is modeled by 
	a 
	GP mean function~$\fhkdot$ in~\eqref{eq:fh} which is updated according to 
	the event-triggering law~\eqref{eq:event} and \cref{alg:adaptive_ctrl}. 
	Then, the tracking error~$\e$ is globally asymptotically stable for 
	any~$\kc>0$ and the inter-event time~$\dt_\kap\coloneqq 
	\tkp-\tk$ is lower bounded by a positive constant~$\tlb>0$, for 
	all~$\kap\in\Nset$ with probability~$1-\delta$.
\end{thm}
\begin{IEEEproof}
	We consider again the common Lyapunov candidate~\eqref{eq:V} and its time 
	derivative
	\begin{align}
	\label{eq:dotVgb1usedf}
	\dot{V}_\kap(\x) \leq  r \df_\kap(\x) -  \kc r^2, 
	\end{align}
	where~$\df_\kap$ is the model error defined in~\eqref{eq:def_df}.
	With noiseless measurements, a GP mean function passes through each 
	training point~\cite{umlauft2018scenario}. Thus, the estimate~$\fhk$ is 
	exact 
	for the time step~$\tk$, and~$\dot{V}_\kap(\x(\tk)) = - \kc r^2$. 
	For~$\tk<t<\tkp$ the estimation error~$\df_\kap(\x(t))$ 
	continuously changes and is generally larger than zero. But
	the term~$\kc r^2$ will dominate~$r \df_\kap(\x)$ with 
	probability~$1-\delta$ 
	by design of the triggering condition~\eqref{eq:event} and 
	\cref{lem:mtotrue}, thus
	\begin{align}
		\Pr \left\{\dot{V}_\kap(\x)<0, \allxt,\kap \in\Nset_0 \right\} \geq 
		1\! -\! \delta.
	\end{align}	
	From \cref{thm:UUB} it is known that the system reaches a compact 
	set~$\Xsett$ for any initial condition~$\x_0 \in \Xset$. Therefore,
	\cref{lem:mtotrue} is applicable and with the radial boundedness of~$V$, 
	the global	asymptotic stability with probability~$1 -\delta$ is shown. 
	
	To show, that the inter-event time is lower bounded, we define the 
	Lipschitz constant~$\Lsig>0$, such that~$\dot{\sigma}_\kap\leq \Lsig 
	\dot{r}$, 
	which exists due to the differentiability of~$\sigma_\kap$ with respect 
	to~$r$.
	Following the lines of~\cite{tabuada2007event}
	\begin{align*}
	\frac{d}{dt} \left| \frac{\sigma_\kap}{r}\right| 
	&=\frac{d}{dt} \frac{\sqrt{\sigma_\kap^2}}{\sqrt{r^2}} 
	=\frac{\dot{\sigma}_\kap r - \sigma_\kap \dot{r}}{r^2} \\
	&\leq \left|\frac{\dot{\sigma}_\kap}{r}\right| 
		+\left|\frac{\sigma_\kap \dot{r} }{r^2} \right|\\
	&\leq \left|\frac{\Lsig(\df_\kap-\kc r)}{r} \right|
		+\left|\frac{\sigma_\kap(\df_\kap-\kc r) }{r^2}\right|\\
	&\leq \Lsig \left| \frac{\df_\kap}{r}\right| + \Lsig \kc	
		+ \left|\frac{\df_\kap\sigma_\kap}{r^2}\right|
		+ \kc\left|\frac{\sigma_\kap}{r}\right|,
	\end{align*}
	and using \cref{lem:mtotrue} yields
	\begin{align*}
		\Pr\bigg\{\frac{d}{dt} \left| \frac{\sigma_\kap}{r}\right| \leq & \Lsig 
		\beta_\kap \left| \frac{\sigma_\kap}{r}\right| + 	\Lsig \kc	
		+ \beta_\kap\left|\frac{\sigma_\kap}{r}\right|^2	\\ 
		&+ \kc\left|\frac{\sigma_\kap}{r}\right|,\allxt\!,\!k\in\Nset_0  
		\bigg\}\geq 1-\delta
	\end{align*}
	for which we define~$\phi= \left|\frac{\sigma_\kap}{r}\right|$. The 
	differential 
	equation
	\begin{align}
	\label{eq:phiode}
		\dot{\phi}= \beta_\kap\phi^2  + \phi(\Lsig \beta_\kap+\kc) + \Lsig\kc,
	\end{align}
	with initial condition~$\phi(\tk)=0$ (from~$\sigma_\kap(\x(\tk))=0$) 
	yields
	\begin{align}
		\label{eq:phisol}
		\phi(t) = \frac{1}{2\beta_\kap}\left(c_1 
		\tan\left(\frac{1}{2}\left((t-\tk) c_1 \pm c_2 \right)\right) 
		-\Lsig \beta_\kap-\kc\right),
	\end{align}
	according to~\cite{wolframalpha2018dgl} for the time interval~$t\in[\tk\ 
	\tkp]$ where~\mbox{$c_1 = \sqrt{4\beta_\kap\Lsig\kc-(\Lsig 
	\beta_\kap+\kc)^2}$}
	and~\mbox{$c_2 = 
	2\arccos\left(\frac{-c_1}{2\sqrt{\beta_\kap\Lsig\kc}}\right)$}. 
	By design, the event is triggered at~$\phi = \kc/\beta_\kap$, which leads 
	to the lower bound on the inter-event time of
	\begin{align*}
		\dt_\kap 
		&\geq \Big(2\arctan\big((3\kc+\Lsig\beta_\kap)/c_1\big)+c_2\Big)/c_1 \\
		&\geq (\pi +c_2)/c_1 \eqqcolon \tlb,
	\end{align*}	
	where $\arctan(\xi)<\pi/2$,~$\forall \xi>0$ is used.
\end{IEEEproof}
Alternatively we consider the scenario, that the model error can continuously 
be monitored.
\begin{assum}
	\label{assum:contmeasure}
	Measurements of~$\x,\dot{x}_n$ are continuously available without effort.
\end{assum}
To take advantage of this assumption, we propose the following event-trigger
\begin{align}
	\label{eq:event_nonProb}
	\tkp \coloneqq \left\{t>\tk\left| \df_\kap(\x)\geq\kc |r|\right. \right\},
\end{align}
which allows to drop \cref{assum:RKHSnorm} and the probabilistic nature of 
\cref{thm:asymEvent} ($\delta = 0$) as formalized in the following.
\begin{cor}
	\label{cor:asymEvent_nonProb}
	Consider the system~\eqref{eq:sys} and a desired trajectory~$\x_d(t)$ under 
	\cref{assum:fgbounded,assum:TrainingData,assum:degree_signg,assum:x_d,assum:perfectg,assum:noisefree,assum:contmeasure}.
	Further consider the control law~\eqref{eq:uf}, where~$\fdot$ is modeled by 
	a GP mean function~$\fhkdot$ in~\eqref{eq:fh} which is updated according to 
	the event-triggering law~\eqref{eq:event_nonProb} and 
	\cref{alg:adaptive_ctrl}. 
	Then, the tracking error~$\e$ is globally asymptotically stable for 
	any~$\kc>0$ and the inter-event time~$\dt_\kap\coloneqq 
	\tkp-\tk$ is lower bounded by a positive constant~$\tlb>0$.
\end{cor}
\begin{IEEEproof}
	This is a direct consequence of the proof for \cref{thm:asymEvent}.
\end{IEEEproof}
We want to highlight, that this requires measurements at any continuous time 
instance, which is generally not possible due to nonzero update rates of 
digital sensors. Therefore, \cref{cor:asymEvent_nonProb} is rather stated for 
completeness. Nevertheless, the algorithm remains data-efficient despite the 
infinity measurements in finite time, because data points are only stored if 
actually needed.

\subsection{Ultimate boundedness for noisy measurements}
\label{sec:noisy}
In case of noisy measurements of~$\dot{x}_n$ (\cref{assum:noisefree} does not 
hold), it is possible to find an ultimate bound to which the system converges. 
The difference to \cref{thm:compB} is that we now make use of the 
event-triggered model update (\cref{thm:compB} allowed arbitrary updates), 
which shrinks down the size of the ultimate bound to a size which is 
proportional to the noise level. From \cref{thm:asymEvent} we derive the 
following results.
\begin{cor}
	\label{cor:UUBEvent}
	Consider the system~\eqref{eq:sys} and a desired trajectory~$\x_d(t)$ under 
	\cref{assum:fgbounded,assum:TrainingData,assum:degree_signg,assum:x_d,assum:perfectg,assum:RKHSnorm}.
	Further consider the control law~\eqref{eq:uf}, where~$\fdot$ is modeled by 
	a GP mean function~$\fhkdot$ in~\eqref{eq:fh} which is updated according to 
	the event-triggering law
	\begin{align}
		\label{eq:eventUUB}
		\tkp \coloneqq \left\{t>\tk\left| \beta_\kap\sigma_\kap(\x)\geq\kc |r| 
		\cap \e \notin \Bset_{\sigonsd} \right. \right\}.
	\end{align}
	and \cref{alg:adaptive_ctrl} where
	\begin{align}
	\Bset_{\sigonsd} = \left\{\e\in \Xsett \left| \norm{\e}
	\leq\frac{\sigonsd\beta_\kap}{\kc\left\| \begin{bmatrix} \ckb\T & 1 
		\end{bmatrix}\right\|}\right.\right\}.
	\end{align}
	Then, the tracking error~$\e$ is GUUB to the set~$\Bset_{\sigonsd}$ for 
	any~$\kc>0$ and the inter-event time~$\dt_\kap$ is lower 
	bounded by a 
	positive constant~$\tlb'>0$, for all~$\kap\in\Nset$ with 
	probability~$1-\delta$.
\end{cor}
\begin{IEEEproof}
	In contrast to \cref{thm:asymEvent}, a measurement at time~$\tk$ does not 
	lead to~$\df_\kap(\x(\tk))=0$, but we make use of the fact that the 
	variance function of a GP~\eqref{eq:s2GP} at any training 
	point can be upper bounded in terms of the measurement noise. Considering 
	the variance for a single training data point at~$\x(\tk)$ as an upper 
	bound for the variance function (which holds according 
	to~\cite{umlauft2017feedback}), the following is concluded
	\begin{align}
		\sigma_\kap(\x(\tk))
		\leq \sqrt{\sigma_{\f}^2-\frac{\sigma_{\f}^4}{\sigma_{\f}^2+\sigonsd^2}}
		= \sqrt{\frac{\sigonsd^2}{1+\sigonsd^2/\sigma_{\f}^2}} 
		< \sigonsd
	\end{align}
	for~$\sigma_{\f}^2<\infty$ using~$k_{\f}(\x,\x) = \sigma_{\f}^2$ 
	in~\eqref{eq:s2GP}.
	Considering again the Lyapunov function~\eqref{eq:V} and its time 
	derivative
	\begin{align}
		\dot{V}_\kap(\x(\tk))\leq  |r|(\beta_\kap 
		\sigonsd- \kc |r|),
	\end{align}
	it is clear that inside of~$\Bset_{\sigonsd}$ negative definiteness of 
	$\dot{V}_\kap$ cannot not be ensured. But outside of this ball it is 
	negative definite and therefore GUUB can be shown similarly to 
	\cref{thm:asymEvent}.
	
	To exclude Zeno behavior only~$ \e \notin \Bset_{\sigonsd} $ must be 
	analyzed, since inside~$\Bset_{\sigonsd}$ no events are triggered.
	The lower bound on the inter-event time is derived along the lines of 
	\cref{thm:asymEvent}. Hence, the dynamics of~$\phi(t)$ as derived 
	in~\eqref{eq:phiode} are the same for the noisy case, but the initial 
	condition~$\phi(\tk)$ is now unequal from 
	zero (due to the noise). However, it can be upper bounded by
	\begin{align*}
		\phi(\tk) <\sqrt{\frac{\sigonsd^2}{1+\sigonsd^2/\sigma_{\f}^2}} /|r| 
		\coloneqq \phi_0.
	\end{align*}
	The solution for the zero initial condition in~\eqref{eq:phisol} is adapted 
	to a nonzero initial condition~$\phi_0$ according 
	to~\cite{wolframalpha2018dgl} by changing~$c_2$ to
	\begin{align}
		\label{eq:phisolnonzeroinit}
		c_2'=2\arctan \left(\frac{2 \beta_\kap \phi_0 +\Lsig 
		\beta_\kap+\kc}{c_1} \right).
	\end{align}
		Accordingly, the lower bound on the inter event time is 
			\begin{align*}
		\dt_\kap \geq (\pi +c_2')/c_1 \eqqcolon \tlb',
		\end{align*}
		which concludes the proof.
\end{IEEEproof}

\subsection{Forgetting strategies}
With the event in \eqref{eq:event} and \cref{alg:adaptive_ctrl}, we 
have 
proposed a strategy, which adds data points to the dataset only if necessary to 
ensure further convergence of the system. However, this still leads to a 
growing computational burden for computing the GP model as the cardinality of 
the dataset~$\Dk$ monotonically increases with time. Particularly, if the 
desired 
trajectory covers a large area in the state space or when high precision 
tracking is 
required, keeping up the real-time capability of the adaptation algorithm is 
challenging. A common technique to circumvent this problem is a forgetting 
mechanism 
(deleting old data points when new ones are added) if a particular budget is 
reached. While most other works, e.g.~\cite{chowdhary2015bayesian}, use a 
heuristic for the forgetting strategy, we propose a \textit{safe} forgetting 
rule, which requires to store only a single data point.
\begin{cor}
	\label{cor:forgetall}
	Consider the system~\eqref{eq:sys} and a desired trajectory~$\x_d(t)$ under 
	\cref{assum:fgbounded,assum:TrainingData,assum:degree_signg,assum:x_d,assum:perfectg,assum:RKHSnorm,assum:noisefree}.
	Further consider the control law~\eqref{eq:uf}, where~$\fdot$ is modeled by 
	a GP mean function~$\fhkdot$ in~\eqref{eq:fh}. This estimate is updated at 
	the event~$\tkp$ in~\eqref{eq:event}, where at each event~$\kap$ 
	all old data points are eliminated from the dataset, thus
	\begin{align*}
		\Dk = \left\{\x(\tk),\dot{x}_n(\tk)\right\}.
	\end{align*}
	Then, with probability~$1-\delta$, the tracking error~$\e$ is globally 
	asymptotically stable for any~$\kc>0$. 
\end{cor}
\begin{IEEEproof}
	This follows along the lines of the proof of \cref{thm:asymEvent}. With the 
	continuity of~$\sigma(\x)$, which is zero at the single training 
	point,~$\sigma(\x(\tk))=0$, it follows that there exists a 
	neighborhood of~$\x(\tk)$ for which~$\sigma_\kap(\x)<\kc |r|$ holds. Thus 
	the results from \cref{cor:forget,thm:asymEvent} are applicable. 
\end{IEEEproof}
Deleting all old data points is consequent in terms of data efficiency, but in 
general triggers events more frequently. Thus, by storing more than one data 
point future measurements can be avoided particularly for periodic desired 
trajectories. For a fixed budget~$\bar{N}\in \Nset$, we can also forget 
unnecessary points and still guarantee stability.
\begin{cor}
	\label{cor:forget}
	Consider the system~\eqref{eq:sys} and a desired trajectory~$\x_d(t)$ under 
	\cref{assum:fgbounded,assum:TrainingData,assum:degree_signg,assum:x_d,assum:perfectg,assum:RKHSnorm,assum:noisefree}.
	Further consider the control law~\eqref{eq:uf}, where~$\fdot$ is modeled by 
	a GP mean function~$\fhkdot$ in~\eqref{eq:fh}. 
	This estimate is updated at the event~$\tkp$ in~\eqref{eq:event}, where at 
	each event~$\kap$ the dataset~$\Dk$ is limited to hold at most~$\bar{N}\in 
	\Nset$ 	data points, such that 
	\begin{align}
	\label{eq:condforget}
	\beta^{\text{elim}}_\kap\sigma^{\text{elim}}_\kap(\x)<\kc |r|
	\end{align} 
	remains true, where~$\beta^{\text{elim}}_\kap$ 
	and~$\sigma^{\text{elim}}_\kap$ denote the values after the elimination.
	Then, with probability~$1-\delta$, the tracking error~$\e$ is globally 
	asymptotically stable for any~$\kc>0$. 
\end{cor}
\begin{IEEEproof}
	This follows along the lines of the proof of \cref{thm:asymEvent}. By 
	\cref{cor:forgetall}, it is known, that there always exists a reduced 
	dataset which fulfills~\eqref{eq:condforget} for~$\bar{N}\geq 1$.
\end{IEEEproof}
Finding the reduced dataset is not a trivial combinatorial problem, but we 
refer to the existing literature for efficient algorithms~\cite{chu1998genetic}.
Note, that the reduced dataset must necessarily contain the most recent 
measurement at~$\tk$ as otherwise the event would not have been triggered.

\subsection{Discussion}
\label{sec:discussControl}

From a control perspective, the most important advantage of GPs is the 
quantification of the uncertainty, i.e. an upper bound of the model error as 
given in by \cref{lem:mtotrue}. We note that the prerequisite for this lemma, 
the bounded RKHS norm in \cref{assum:RKHSnorm}, is difficult to verify, however 
minimal assumptions are necessary as otherwise a generalization beyond the 
training data is impossible~\cite{wolpert2002supervised}. 

Also the maximum mutual information~$\gamma_\kap$ in \cref{lem:mtotrue} cannot 
be computed analytically for a general kernel, but we refer to the existing 
literature~\cite{srinivas2012information}, which provides upper bounds 
on~$\gamma_\kap$ for different kernels (including the squared exponential 
kernel). Since~$\beta_\kap$ is not 
trivial to find, we would like to point out that~$\beta_\kap$ always appears 
in the ratio with~$\kc$, thus any conservatism/approximation in~$\beta_\kap$ 
can be compensated generally by the designers choice of the control gain~$\kc$.

Overcoming these challenges, the GP allows - based on event-triggered online 
learning - to design a feedback linearizing control law, which asymptotically 
stabilizes an 
initial unknown system (with high probability). This is made possible by the 
error bounds on the model which is the most significant advantage of a GP over 
alternative modeling approaches like neural networks~\cite{lewis1998neural}.

As the model update is event-triggered, only data points which are necessary 
to increase the precision of the model are collected. This reduces the 
frequency at which measurements are taken and increases data efficiency. With 
\cref{cor:forgetall} we have shown, that only a single data point must be 
stored to guarantee asymptotic convergence. This is a significant 
advantage of the locally linearizing control law in comparison to predictive 
control laws or reinforcement learning algorithms, where an accurate global 
model is required. Accurate global models require active exploration, e.g. 
through exploration noise, which sacrifices control performance (known as 
exploration-exploitation trade-off).

\section{Numerical Illustration}
\label{sec:sim}

To illustrate the proposed approach, we present simulations\footnote{The code 
is available at \url{https://gitlab.lrz.de/ga68car/adaptFeLi4GPs}} for the 
control affine system
\begin{align}
	\label{eq:simsys}
	\dot{x}_1 &= x_2, \\
	\dot{x}_2 &= \underbrace{1-\sin(x_1) + 	s(x_2)}_{=\fx}
		+\underbrace{ \left(1+\frac{1}{2}\sin(x_2/2)\right)}_{=\gx} u,
			\nonumber
\end{align}
where~$s(x_2) = \frac{0.5}{1 + \exp(-x_2/10)}$ is the sigmoidal function. It is 
a modified pendulum system and fulfills 
\cref{assum:fgbounded,assum:degree_signg}. To 
ensure \cref{assum:RKHSnorm} holds, we do not simulate directly 
on~\eqref{eq:simsys} but use a GP mean which was trained on it with a high 
density of training points. 
 As we are working in simulation 
 \cref{assum:perfectg,assum:TrainingData,assum:noisefree} do hold 
 or do not hold by design in the following two different scenarios, which we 
 use to illustrate the proposed approach. An overview of the employed 
parameters is given in \cref{tab:params}

\subsection{Scenario 1: Time-triggered updates}
\label{sec:simSec1}
In Scenario 1 (S1), we illustrate the results from \cref{sec:control} 
which are shown to hold for an arbitrary switching sequence. Therefore, we 
utilize a periodic, time-triggered model update, thus~$\tkp-\tk = 
\dt$,~$\forall \kap$, 
with~$\dt=0.5$ and follow \cref{alg:adaptive_ctrl}.
We consider~$\fx$ and~$\gx$ to be unknown, so \cref{assum:perfectg} does not 
hold, but we know that~$\gx$ is positive, so \cref{assum:degree_signg} holds. 
As 
reference trajectory 
\begin{align}
	x_d(t) = 1-\frac{1}{1+\exp(-20(t-10))}
\end{align}
is used, which describes a ``soft'' jump from~$x_1=1$ to~$x_1=0$ 
at~\mbox{$t=10$} and 
it fulfills the required smoothness in \cref{assum:x_d}.
The scenario works on noisy measurements, thus \cref{assum:noisefree} does not 
hold.
As this scenario does not utilize \cref{assum:RKHSnorm}, we consider the 
kernel's hyperparameters to be unknown. Therefore, an hyperparameter 
optimization according to~\eqref{eq:opt_like} is performed at each model update 
step~$\kap$.
The simulation is stopped manually after~$T_{\text{sim}}=20$, which leads to 
$N=40$ data points. 

\begin{figure}
	\centering 
	\includegraphics[width = 0.48\textwidth]{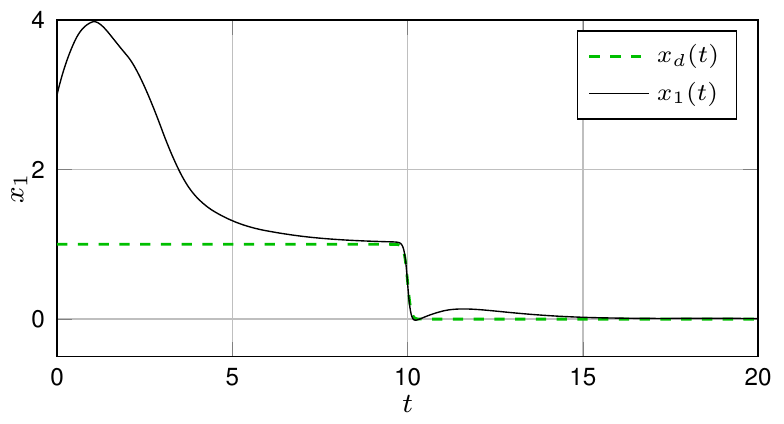}
	\caption{Scenario 1: The black solid line illustrates the actual, the green 
		dashed line 
		the desired value for the state~$x_1$. The system converges to the 
		desired state over time.}
	\label{fig:fgtime_time}
\end{figure}

\begin{figure}
	\centering 
		\includegraphics[width = 0.48\textwidth]{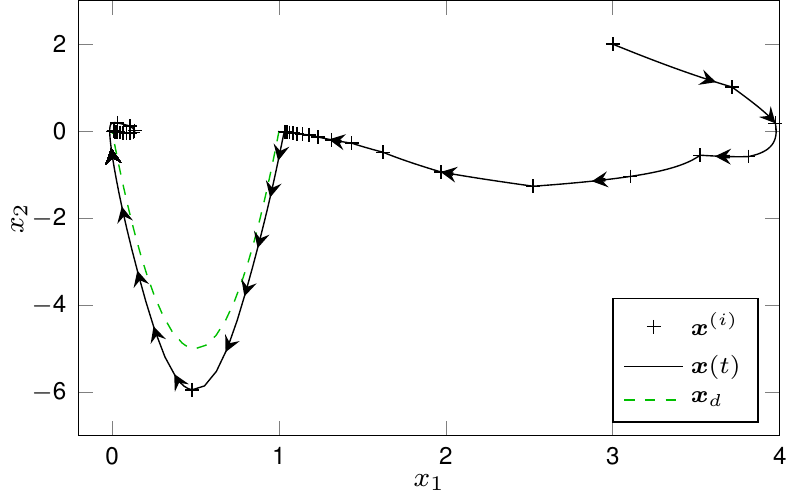}
	\caption{Scenario 1: Black crosses indicate the collected training points, 
	the black 	
			solid line illustrates the actual, the green dashed the desired 
			trajectory. The system approaches the desired states as more 
			training points are collected.}
	\label{fig:fgtime_ss}
\end{figure}

\begin{figure}
	\centering 
			\includegraphics[width = 0.48\textwidth]{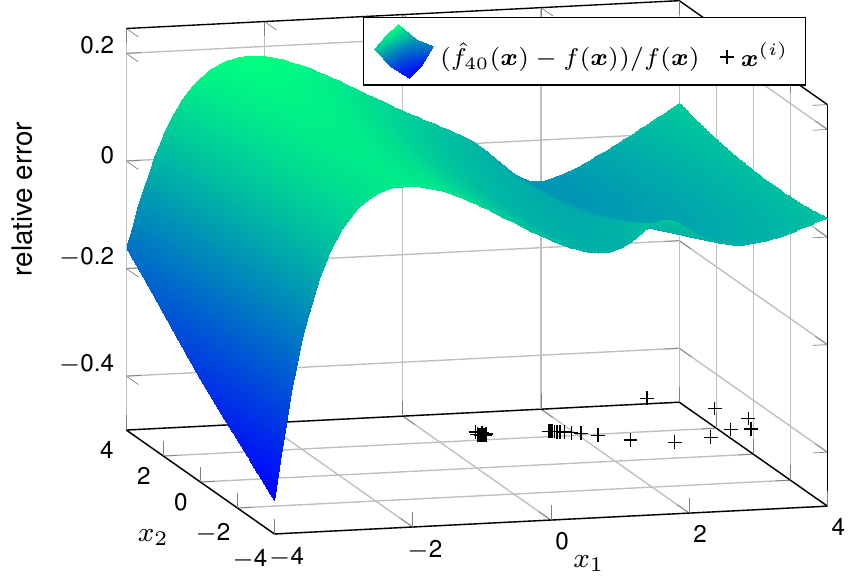}
	\caption{Scenario 1: The surface illustrates the relative error between the
			 true function~$\fx$ and the model estimate~$\fh_{40}(\x)$ after 
			 taking~$40$ training points (black crosses). The error is the 
			 lowest 
			 (in terms of absolute value) near the training data.
		} 
	\label{fig:fgtime_fvsfh}
\end{figure}

\begin{figure}
	\centering 
	\includegraphics[width = 0.48\textwidth]{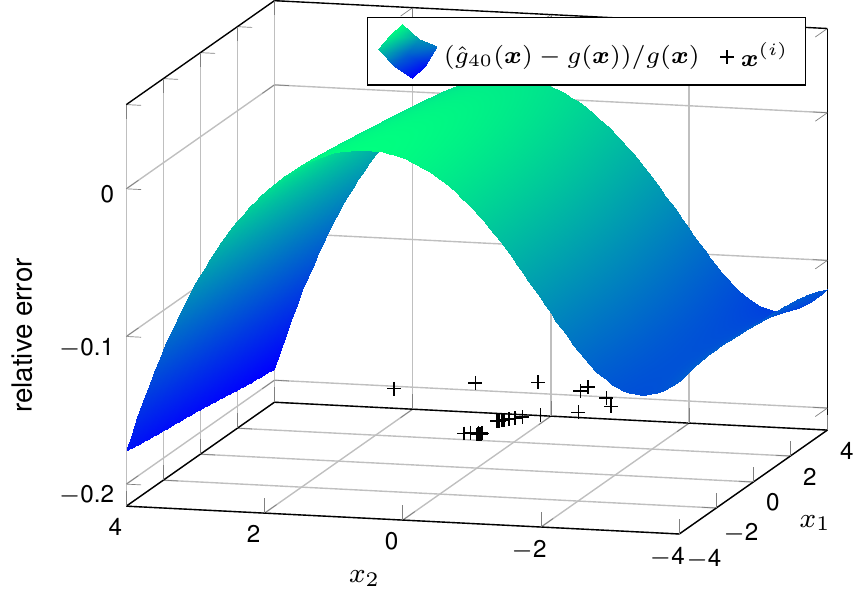}
	\caption{Scenario 1: The surface illustrates the relative error between the 
			true function~$\gx$ and the model estimate~$\gh_{40}(\x)$ after 
			taking~$40$ training points (black crosses). The error is the 
			lowest 
			(in terms of absolute value) near the training data.
		} 
	\label{fig:fgtime_gvsgh}
\end{figure}

\Cref{fig:fgtime_time} shows the desired trajectory and the corresponding 
tracking performance of the controller over time.
\Cref{fig:fgtime_ss} illustrates the resulting trajectory in the state space.
In \cref{fig:fgtime_fvsfh,fig:fgtime_gvsgh} the true system dynamics, 
$\fx,\gx$ are compared with the approximations~$\fhx,\ghx$ at  
the end of the simulation.
It turns out, that the hyperparameters for~$k_{fg}$ are well identified: 
With~$l_{1,f}\ll l_{2,f}$, the estimate for~$\fhx$ shows that~$\fx$ 
mainly depends on~$x_1$  (and vice versa for~$\gx$ 
with~$l_{1,g}\gg l_{2,g}$).
As it can be seen in \cref{fig:fgtime_fvsfh,fig:fgtime_gvsgh} the estimates are 
more precise near the training data.

It can be seen, that the two stationary points of the desired 
trajectory,~$(1,0)$ and~$(0,0)$, are approached with high 
precision in the steady state. However, for both it requires a few measurements 
to be collected in the corresponding  area of the state space and the following 
model updates to achieve this high precision. Once reached the steady state, 
the time-triggered implementation keeps adding unnecessary data points even 
though the model already has a high precision in this area. This is improved 
with the event-triggered model update as illustrated in Scenario 2.

\begin{table}
	\centering \vspace{1.5mm}
	\begin{tabularx}{8.6cm}{c c c c c c c c c}
		$\kc$ & $\lambda$ & $m_g(\x)$&$\sigon$(S1)&$\sigon$(S2)& $\x_0$ &  
		$\beta$& $r_{\text{min}}$	\\ \hline \rule{0pt}{3ex}   
		$1$   & $1$ & $=2$, $\forall \x$ 	 & $10^{-6}$& $10^{-16}$ &$[3 
		\ 2]\T$ & 7 & $10^{-5}$ 
		\rule{0pt}{3ex}   
	\end{tabularx}
	\caption{Simulation parameters}
	\label{tab:params}
	\vspace{-0.3cm}
\end{table}

\subsection{Scenario 2: Event-triggered updates}
\label{sec:simSec2}
In Scenario 2 (S2), the results in \cref{sec:event} are illustrated, which 
utilizes the event-triggered model update described by \eqref{eq:event}. For 
this scenario,~$\gx$ is assumed to be known (\cref{assum:perfectg} holds). The 
reference trajectory
\begin{align}
	x_d(t) = \sin(t)
\end{align}
is used, which describes a circle with radius~$1$ in the state space. The 
scenario works on noise free measurements, thus \cref{assum:noisefree} does 
hold, however, for numerical stability a minimal noise is assumed 
($\sigon=10^{-16}$).
The simulation is stopped manually after~$T_{\text{sim}}=100$.

As this scenario utilizes \cref{assum:RKHSnorm}, we take the kernel 
hyperparameters to be known at~$\sigma_{\f}^2=5, l_{1/2,\f}^2=5$ 
and do not update these at any of the triggered events. Additionally, we 
set~$\beta_\kap$ constant $\forall_\kap$ and refer to the discussion in 
\cref{sec:discussControl} and~\cite{berkenkamp2016safe}.
Additionally, we enforce a lower bound~$r>r_{\text{min}}$ to 
avoid numerical difficulties.

\begin{figure}
		\centering 
	\includegraphics[width = 0.48\textwidth]{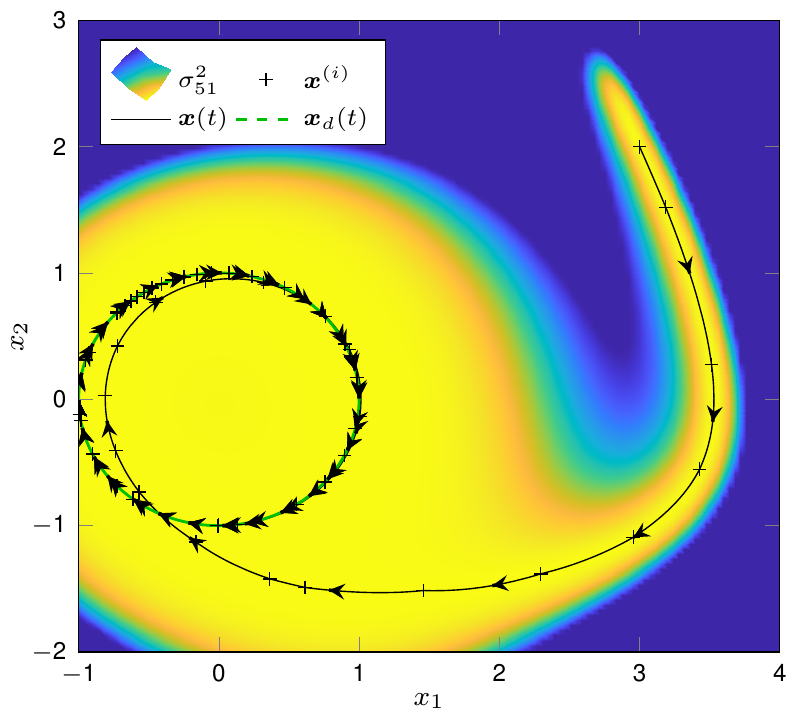}
	\caption{Scenario 2: Black crosses indicate the collected training points, 
	the black 	
		solid line illustrates the actual, the green dashed the desired 
		trajectory. The colormap shows the variance function~\eqref{eq:s2GP} 
		for the GP~$\sigma_{51}(\x)$ after the 51st update, where yellow 
		indicates low variance and blue high variance. }
	\label{fig:fevent_ss}
\end{figure}

\begin{figure*}
	\centering 
			\includegraphics[width = \textwidth]{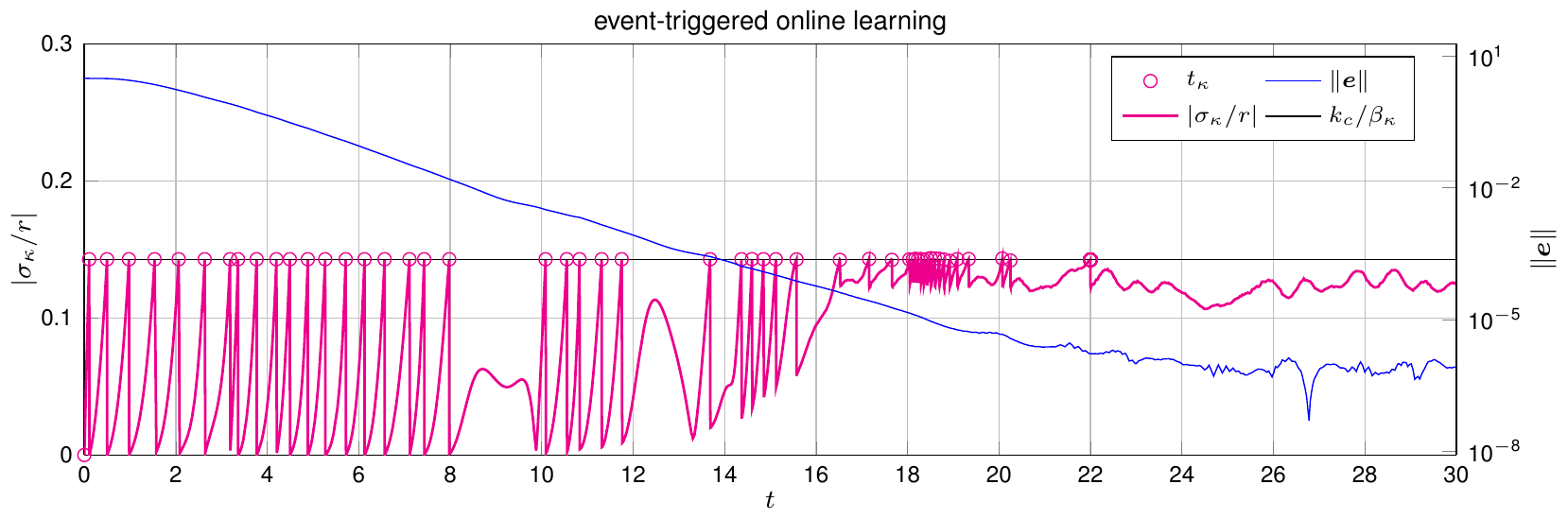}	
			\includegraphics[width = \textwidth]{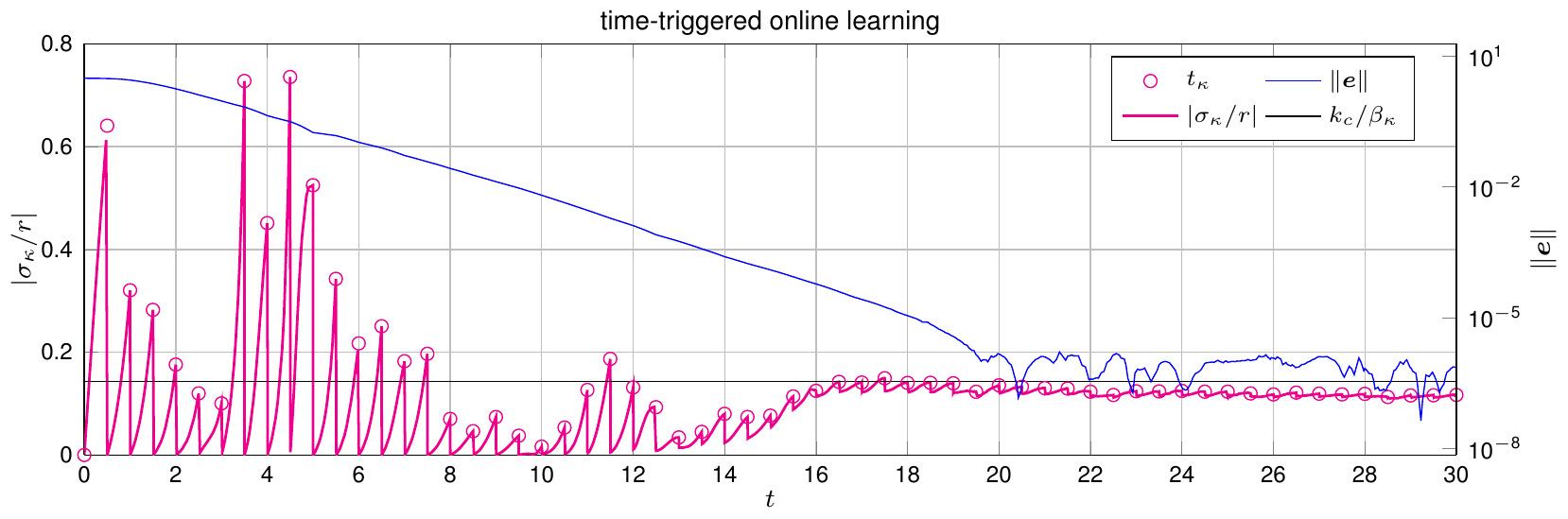}	
	\caption{Scenario 2: Comparison of the event-triggered (top) and 	
			time-triggered (bottom) online learning. For the first, events 
			(magenta circles) are triggered when the threshold~$\kc/\beta_\kap$ 
			(black vertical line) is reached by~$\sigma_\kap/r$ as proposed 
			in~\eqref{eq:event}. For the latter, events are triggered after a 
			fixed time interval ($\dt=0.5$). The blue lines show the norm of 
			the tracking error~$\norm{\e}$.}
	\label{fig:fevent_time}
\end{figure*}

\Cref{fig:fevent_time} shows the tracking error until~$t=30$, which decreases 
initially approximately exponentially until a numerical limit is reached. In 
the event-triggered setup, a total of~$51$ events are triggered until 
sufficient training points are collected around the 
desired trajectory. This is also visualized in the state space view in 
\cref{fig:fevent_ss}. In comparison, the time-triggered approach requires to 
store~$200$ data points and would keep adding points for 
longer simulations, which the event-based would not.

The time-trigger results in a higher computational burden of the data-driven 
approach: The event-triggered simulation only takes~$\approx 5s$ on a Matlab 
MATLAB 2019a implementation on a i5-6200U CPU with 2.3GHz and uses $0.20957$MB. 
The time-triggered approach take~$\approx 11s$ and uses~$0.47924$MB of memory.

\Cref{fig:fevent_time} also shows, that the stability criteria in 
\cref{thm:asymEvent} is fulfilled for the event-triggered case, 
since~$\beta_\kap\sigma_\kap(\x)\leq\kc |r|$ holds for any time. In contrast - 
for the time-triggered case - this condition is violated frequently, which 
means that negative definiteness of the common Lyapunov function - and thereby 
stability - cannot be shown.

\section{CONCLUSION} 
\label{sec:conclusion}

This article proposes an online learning feedback linearizing control law based 
on Gaussian process models. The closed-loop identification of the initially 
unknown system exploits the control affine structure by utilizing a composite 
kernel. The model is updated event-triggered, taking advantage of the 
uncertainty measure of the GP. The control law results in global asymptotic 
stability of the tracking error in the noiseless case and in global uniform 
ultimate boundedness for noisy measurements (of the highest state derivative) 
with high probability. We therefore propose a safe and data-efficient online 
learning control approach because model updates occur only if required to 
ensure stability. Zeno behavior is excluded as a lower bound on the inter-event 
time is derived. 
The proposed techniques are illustrated 
using simulations to support the theoretical results.

\appendices
\section{Expressing structure in kernels}
\label{sec:kernels}
According to~\cite{duvenaud2014thesis}, the kernel of the GP does not only 
determine the smoothness properties of the resulting functions but can also be 
utilized to express prior knowledge regarding the structure of the unknown 
function.
\subsection{Sum of functions}
\label{sec:GPSum}
Consider \mbox{$f_a,f_b \colon \Xset \to \Rset$} which both originate from two 
independent GP priors
\begin{align}
f_a(\x) &\sim \GP\left(m_a(\x), k_a(\x,\x')  \right),  \\
f_b(\x) &\sim \GP\left(m_b(\x), k_b(\x,\x')  \right),
\end{align}
and add up to~$\fsum\colon \Xset \to \Rset$, thus~$\fsum(\x) = f_a(\x) + 
f_b(\x)$. Then,
\begin{align}
\fsum(\x) \sim \GP\left(m_a(\x)+m_b(\x), k_a(\x,\x') + k_b(\x,\x') \right)
\end{align}
is also a GP with kernel~$k_a(\x,\x') + k_b(\x,\x')$. 

For regression, where noisy measurements 
with~\mbox{$\epsilon^{(i)}\sim\N(0,\sigon)$} of the sum of the two function are 
available
\begin{align}
\ysum^{(i)} =\fsum\left(\xhiv\right) + \epsilon^{(i)}
=  f_a\left(\xhiv\right) +	f_b\left(\xhiv\right) +	
\epsilon^{(i)},
\end{align}
with~$i=1,\ldots, N$, the joint distribution of the individual functions and 
the 
observations is given by
\begin{align}
\begin{bmatrix}f_a(\x^*)\\ f_b(\x^*)\\ \ybsum\end{bmatrix}\sim\N\left(\vec{0}, 
\begin{bmatrix}
k_a^*  &  \vec{0} & \k_a\T  \\
\vec{0} &  k_b^* & \k_b\T  \\
\k_a  & \k_b & \K_a +\K_b + \sigon \vec{I}_N
\end{bmatrix} \right),
\end{align}
where the prior mean functions are set to zero~$m_a(\x)=m_b(\x)=0$ for 
notational simplicity and~$\k_a,\k_b,k_a^*, k_b^*$ are defined 
according to~\eqref{eq:defk}. 
By conditioning, the output of~$f_a$ and~$f_b$ are inferred for a test 
points~$\x^*$
\begin{align}
\left. f_a(\x^*)\right| \vec{X},\ybsum  &\sim
\N \left( \k_a\T \K_\text{sum}^{-1}\ybsum, 
k_a^* - \k_a\T \K_\text{sum}^{-1}\k_a  \right), \\
\left. f_b(\x^*)\right| \vec{X},\ybsum  &\sim \N \left( \k_b\T 
\K_\text{sum}^{-1}\ybsum, 
k_b^* - \k_b\T \K_\text{sum}^{-1}\k_b  \right),
\end{align}
where~$\K_\text{sum} = \K_a +\K_b + \sigon \vec{I}_N$ with 
$\K_a,\K_b$ according to~\eqref{eq:K}.
Similarly to~\eqref{eq:opt_like}, the extended hyperparameter 
vector~$\bpsi_\text{sum}= [\bpsi_a\T\ \ \bpsi_b\T]\T$ is obtained through 
optimization of the likelihood, where~$\K=\K_\text{sum}$ and~$\y_f 
=\ybsum$. This allows to predict a value of the individual functions~$f_a,f_b$ 
even though only their sum has been measured. 

\subsection{Product with known function}
\label{sec:GPProduct}
Consider an unknown function~$f_h(\x)\colon \Xset \to \Rset$, which is 
multiplied with the known function~$h(\x)\colon \Xset \to \Rset$, we can 
model~$f_h$ using a GP with a scaled kernel function and noisy measurements 
\begin{align}
\yprod^{(i)}=\fprod \left(\xhiv\right)+\epsilon^{(i)} 
= f_h\left(\xhiv\right) 	
h\left(\xhiv\right)+\epsilon^{(i)}
\end{align}
of the product with~$\epsilon^{(i)}\sim \N(0,\sigon),\ \alli$.
Thus, if~\mbox{$f_h \sim \GP(0,k_h(\x,\x'))$} is a GP, then~$\fprod(\x)$ is 
also a GP with kernel 
\begin{align}
k_\text{prod}(\x,\x') = h(\x)k_h(\x,\x')h(\x'),
\end{align}
where the prior mean is set to zero~$m_h(\x) = 0$ for notational simplicity.

The joint distribution of the measurements and the inferred output of~$\fprod$ 
at a test input~$\x^*$ is given by
\begin{align}
\begin{bmatrix} f_h(\x^*)\\ \ybprod \end{bmatrix}
\sim\N\left(\vec{0},\begin{bmatrix}
k_h^* & \k_h\T\vec{H}\T  \\
\vec{H}\k_h  &\vec{H}\T \K_h\vec{H}
+\sigon \vec{I}_N
\end{bmatrix} \right),
\end{align}
where~$\vec{H} = \diag\left(h\left(\x^{(1)}\right), \dots, 
h\left(\x^{(N)}\right)\right)\in \Rset^{N\times N}$ and~$\k_h,\ k_h^*$,~$\K_h$ 
are defined similarly to~\eqref{eq:defk} and~\eqref{eq:K}, respectively.  
By conditioning on the training data and the input the function~$f_h$ is 
inferred by
\begin{align}
\left. f_h(\x^*)\right|\vec{X}, \ybprod  \sim
\N \big( \k_h\T\vec{H}\T& \K_\text{prod}^{\text{-}1}\ybprod, \\
& k_h^* -\k_h\T\vec{H}\T\K_\text{prod}^{\text{-}1}\vec{H}\k_h 
\big), \nonumber
\end{align}
where~$\K_\text{prod} =\vec{H}\T \K_h\vec{H} + \sigon \vec{I}_N$.
\begin{remark}
	Instead of scaling the kernel, it seems more straight forward to 
	use~$\yprod^{(i)}/h\left(\xhiv\right)$ as training data for a GP with 
	unscaled kernel. However, this would scale the observation noise 
	undesirably, is numerically 
	not stable and is not compatible with the summation of kernels in 
	Appendix~\ref{sec:GPSum}, which we combine in our identification approach 
	in \cref{sec:GP4ctrlaffinesys}.
\end{remark}

\section{Improving identification}
\label{sec:improveid}
From \cref{lem:posgh} it is known, that the state remains bounded for any 
finite~$0<T<\infty$ without any control input. Thus, without risking damage to 
the system, one can set~$u=0$ for at time interval~$T$ and record an open-loop 
training point
\begin{align}
y^{(i_{\text{ol}})} = 
\f\left(\x^{(i_\text{ol})}\right)+\epsilon^{(i_\text{ol})}, 
\end{align}
which is highly beneficial as it only measures~$\fx$ (with the usual 
noise~$\epsilon$). The GP framework allows to merge these 
$i_{\text{ol}}=1,\ldots,N_{\text{ol}}$ observations with the 
closed-loop training points in~$\Dk$ to improve the prediction as 
follows: 
Consider the extension of the joint distribution~\eqref{eq:jointfg} (where 
$u=1$ is assumed in the close loop measurements and~$m_g(\x)=0$ for notational 
convenience)
\begin{align*}
\begin{bmatrix} f(\x^*) \\ g(\x^*) \\ \y \\ \y_{\text{ol}} \end{bmatrix} \sim 
\N \left(\vec{0},
\begin{bmatrix} k_f^* & 0 & \k_f\T & \k_{f,\text{ol}}\T \\
0 	  & k_g^* &  \k_g\T   & \vec{0}_{1\times N_{\text{ol}}}\\
\k_f & \k_g  & \K_{fg} & \K_{\text{ol,cl}}\T \\
\k_{f,\text{ol}} & \vec{0}_{N_{\text{ol}}\times 1} & 
\K_{\text{ol,cl}}&\K_{\text{ol}}
\end{bmatrix} \right),
\end{align*}
where~$\K_{\text{ol,cl}}$,~$\K_{\text{ol}}$ are the pairwise evaluation 
of~$k(\x^{(i_\text{ol})},\xhiv)$,
$k\left(\x^{(i_\text{ol})},\x^{(i'_\text{ol})}\right)$ 
and~$\k_{f,\text{ol}}$ evaluates~$k\left(\x^*,\x^{(i_\text{ol})}\right)$.
for all~$\alli$,~$i_{\text{ol}},i'_{\text{ol}}=1,\ldots,N_{\text{ol}}$. Then, 
the estimates are given by
\begin{align}
\fh(\x^*)&= [\k\T_f\ \k\T_{f,\text{ol}}]\vec{\mathcal{K}}^{-1} \tilde{\y},\quad 
\gh(\x^*)=[\k\T_g\ \vec{0}_{1\times N_{\text{ol}}}]\vec{\mathcal{K}}^{-1} 
\tilde{\y} \nonumber\\
\text{with } \vec{\mathcal{K}}&= \begin{bmatrix}
\K_{fg} &\K_{\text{ol,cl}}\T \\
\K_{\text{ol,cl}}&\K_{\text{ol}}
\end{bmatrix},\text{ and } \tilde{\y} = \begin{bmatrix}\y \\ \y_{\text{ol}}
\end{bmatrix}.
\end{align}
We do not further investigate this extension, since it does not provide any 
additional formal guarantees regarding the convergence, however, in practice, a 
significant improvement of the identification can be expected.

\IEEEtriggeratref{48}
\bibliographystyle{IEEEtran}
\bibliography{IEEEabrv,../../literature/root}
\enlargethispage{-1.4cm}
\vspace{-0.8cm}
\begin{IEEEbiography}[{\includegraphics[width=1in,height=1.25in,clip,keepaspectratio]
			{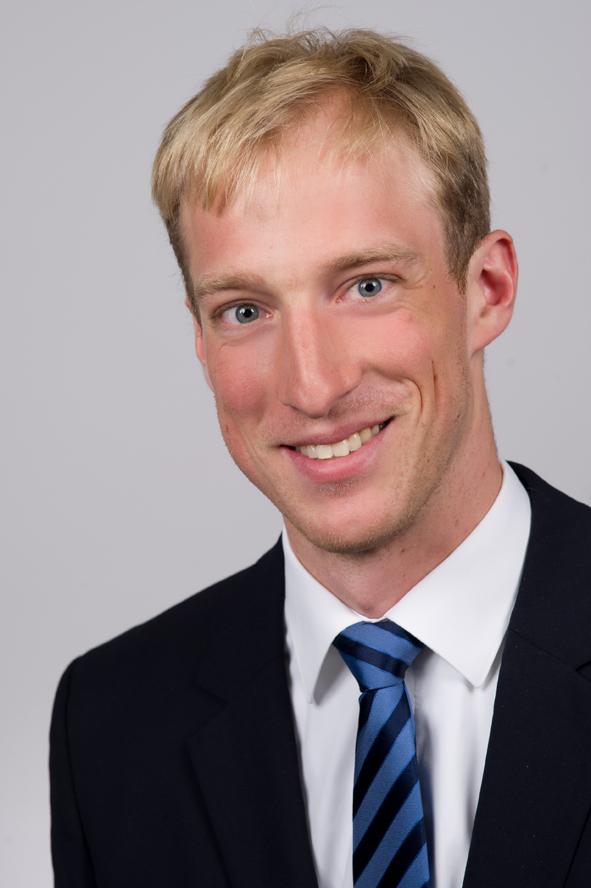}}]{\color{diffc}Jonas Umlauft (S'15)}
				\color{diffc}
	received his B.Sc. and M.Sc. degree in electrical engineering and 
	information technology from the Technical University of Munich, Germany, in 
	2013 and 2015, respectively. His Master’s thesis was carried out at the 
	Computational and Biological Learning Group at the University of Cambridge, 
	UK. Since May 2015, he is a PhD student at the Chair of 
	Information-oriented Control, Department of Electrical and Computer 
	Engineering at the  Technical University of Munich, Germany. 
	His current research interests includes 
	stability of data-driven control systems and system identification based on 
	Gaussian processes. 
\end{IEEEbiography}
\vspace{-0.8cm}
\begin{IEEEbiography}[{\includegraphics[width=1in,height=1.25in,clip,keepaspectratio]
		{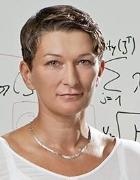}}]{\color{diffc}Sandra Hirche (M’03-SM’11)}
			\color{diffc}
		received the Diplom-Ingenieur degree in aeronautical engineering from 
		Technical University Berlin, Germany, in 2002 and the Doktor-Ingenieur 
		degree in electrical engineering from Technical University Munich, 
		Germany, in 2005. From 2005 to 2007 she was awarded a Postdoc 
		scholarship from the Japanese Society for the Promotion of Science at 
		the Fujita Laboratory, Tokyo Institute of Technology, Tokyo, Japan. 
		From 2008 to 2012 she has been an associate professor at Technical 
		University Munich. Since 2013 she is TUM Liesel Beckmann Distinguished 
		Professor and has the Chair of Information-oriented Control in the 
		Department of Electrical and Computer Engineering at Technical 
		University Munich. Her main research interests include cooperative, 
		distributed and networked control with applications in human-machine 
		interaction, multi-robot systems, and general robotics. She has 
		published more than 150 papers in international journals, books and 
		refereed conferences. Dr. Hirche has served on the Editorial Boards of 
		the IEEE Transactions on Control of Network Systems, IEEE Transactions 
		on Control Systems Technology, and the IEEE Transactions on Haptics. 
		She has received multiple awards such as the Rohde \& Schwarz Award for 
		her PhD thesis, the IFAC World Congress Best Poster Award in 2005 and - 
		together with students - the 2018 Outstanding Student Paper Award of 
		the IEEE Conference on Decision and Control as well as Best Paper 
		Awards of IEEE Worldhaptics and IFAC Conference of Manoeuvring and 
		Control of Marine Craft in 2009.
\end{IEEEbiography}
\end{document}